\documentclass[]{emulateapj}

\def\Mpc{{\rm Mpc}}

\def\kms{{\rm km}\,{\rm s}^{-1}}
\def\yr{{\rm yr}}
\def\Msun{{\rm M}_\odot}
\def\Mstel{M_\ast}
\def\Mdisk{M_{\ast}^{\rm disk}}

\def\Mhalo{M_{\rm halo}}
\def\LCDM{$\Lambda$CDM}
\def\resp{respectively}
\def\bfr{\bf\color{red}}

\def\ssfr{{\rm sSFR}}
\def\sfr{{\rm SFR}}
\def\MS{SFMS}
\def\sigms{\sigma_{\rm \MS}}

\def\Sersic{S\'{e}rsic}

\def\beq{\begin{equation}}
\def\eeq{\end{equation}}
\def\bitem{\begin{itemize}}
\def\eitem{\end{itemize}}
\def\benum{\begin{enumerate}}
\def\eenum{\end{enumerate}}

\mathchardef\mhyphen="2D

\def\cite{{\bfr CITE}}

\usepackage{mathtools}
\usepackage{graphicx}
\usepackage[bottom]{footmisc}
\usepackage{mathptmx}
\usepackage{color}
\usepackage[breaklinks=true, colorlinks=true, linkcolor=red, urlcolor=blue, citecolor=black]{hyperref}
\shortauthors{Abramson et al.}
\shorttitle{[Log-]Normal Galaxy Evolution}

\begin{document}

\title{
Return to [Log-]Normalcy: Rethinking Quenching, The Star Formation Main Sequence, \\ 
And Perhaps Much More
}

\slugcomment{Accepted to The Astrophysical Journal, 1 September 2016}

\author{
Louis E.\ Abramson\altaffilmark{1},
Michael D.\ Gladders\altaffilmark{2},
Alan Dressler\altaffilmark{3}, 
Augustus Oemler, Jr.\altaffilmark{3},
Bianca Poggianti\altaffilmark{4},\\
and Benedetta Vulcani\altaffilmark{5}
}


\begin{abstract}

Knowledge of galaxy evolution rests on cross-sectional observations of different objects at different times. {\it Understanding} of galaxy evolution rests on longitudinal interpretations of how these data relate to individual objects moving through time. The connection between the two is often assumed to be clear, but we use a simple ``physics-free'' model to show that it is not and that exploring its nuances can yield new insights. Comprising nothing more than $2094$ loosely constrained lognormal star formation histories (SFHs), the model faithfully reproduces the following data it was not designed to match: stellar mass functions at $z\leq8$; the slope of the star formation rate/stellar mass relation (the SF ``Main Sequence'') at $z\leq6$; the mean $\ssfr(\equiv\sfr/\Mstel)$ of low-mass galaxies at $z\leq7$; ``fast-'' and ``slow-track'' quenching; downsizing; and a correlation between formation timescale and $\ssfr(\Mstel,t)$ similar to results from simulations that provides a natural connection to bulge growth. We take these findings---which suggest that quenching is the natural downturn of all SFHs affecting galaxies at rates/times correlated with their densities---to mean that: (1) models in which galaxies are {\it diversified} on Hubble timescales by something like initial conditions rival the dominant grow-and-quench framework as good descriptions of the data; or (2) absent spatial information, many metrics of galaxy evolution are too undiscriminating---if not inherently misleading---to confirm a unique explanation. We outline future tests of our model but stress that, even if ultimately incorrect, it illustrates how exploring different paradigms can aid learning and, we hope, more detailed modeling efforts.

\end{abstract}


\altaffiltext{1}{
Department of Physics \& Astronomy, UCLA, 430 Portola Plaza, Los Angeles, CA 90095-1547, USA; \href{mailto:labramson@astro.ucla.edu}{labramson@astro.ucla.edu}
}
\altaffiltext{2}{
Department of Astronomy \& Astrophysics, and Kavli Institute for Cosmological Physics, The University of Chicago, 5640 South Ellis Avenue, Chicago, IL 60637, USA
}
\altaffiltext{3}{
The Observatories of the Carnegie Institution for Science, 813 Santa Barbara Street, Pasadena, CA 91101, USA
}
\altaffiltext{4}{
INAF-Osservatorio Astronomico di Padova, Vicolo Osservatorio 5, 35122 Padova, Italy
}
\altaffiltext{5}{
School of Physics, The University of Melbourne, VIC 3010, Australia
}


\section{Introduction}
\label{sec:intro}

Measurements of galaxy sizes, masses, and star formation rates now span much of cosmic time. Yet, core uncertainties regarding how to interpret these data hinder the construction of a definitive physical narrative of galaxy evolution. This issue lies at the center of a debate between two schools of thought in this field:
\bitem
        \item Galaxy growth via in situ star formation is a uniform phenomenon
        {\it interrupted} by internal or external ``quenching'' processes
        \citep[e.g.,][]{PengLilly10, Steinhardt14}.
        \item Galaxy growth is a heterogeneous phenomenon {\it diversified}
        by even more fundamental processes of which quenching is symptomatic,
        but not necessarily informative. \citep{Tinsley68, Gladders13b, Kelson14}.
\eitem

This distinction is not semantic, but axiomatic: It reflects different ideas about the meaning of the data and ultimately leads to different basic questions in galaxy evolution---``What stops star formation?" or ``What shapes star formation histories?" Hence, it also affects where and how one looks for the key physical mechanisms regulating galaxy growth.

\begin{figure*}[t!]
\centering
\includegraphics[width = 0.67\linewidth, trim = 0.5cm 0.5cm 0.5cm 0cm]{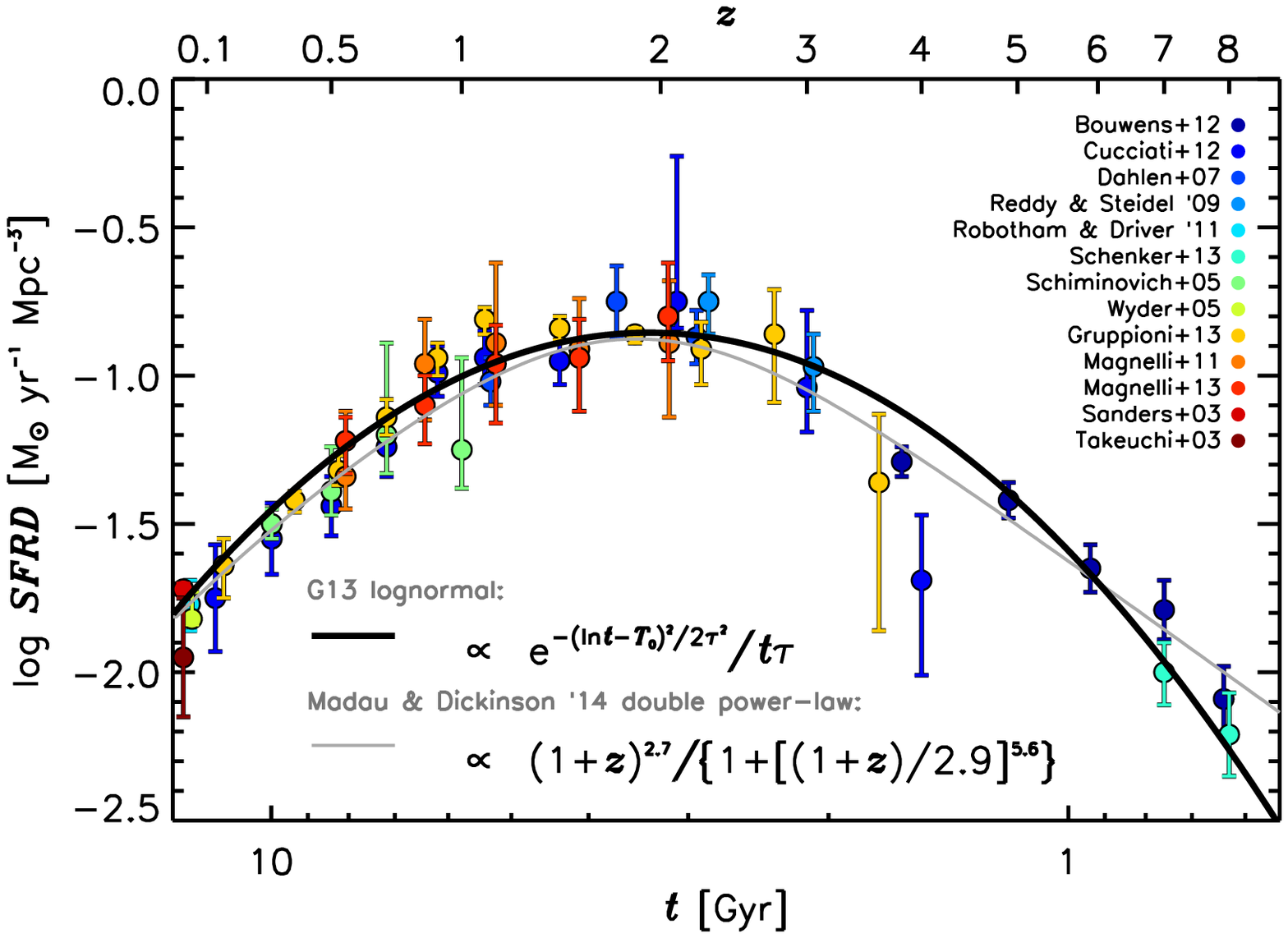}
\caption{The cosmic SFRD$(t)$ is well-described by a simple lognormal in time. Data points are from the studies listed as compiled by \citet{MadauDickinson14}. The thin grey line is the best-fit double power-law from that text. The thick black line is the best-fit lognormal: $(T_{0},\tau)_{\rm Uni} = (1.64, 0.66)$, or (5.16, 1.93) Gyr (Equation \ref{eq:SFRDshape}; $A_{\rm Uni}=0.96~\Msun~{\rm Mpc^{-3}}$). Due to fitting revised data, these values differ immaterially from those in G13 (Appendix \ref{sec:AA}, Figure \ref{fig:manySFRDs}). While the two fits are of similar quality, the lognormal has one fewer free parameter.}
\label{fig:fit}
\end{figure*}

A prolific body of work describes the quenching paradigm and its associated mechanisms, including: starvation from external gas supplies; heating of gas in dark matter halos; various galaxy-galaxy interactions; and feedback from supernovae and/or active galactic nuclei \citep[AGN; e.g.,][]{Larson80, Dekel86, Moore99, Keres05, Keres09b, Springel05c, Croton06, Oppenheimer10, Hopkins14, Hopkins16}.\footnote{``Quenching'' first described abrupt truncations of arbitrary levels of star formation [e.g., (post-)starbursts; \citealt{Harker06}]. The term has broadened to encompass gradual processes that push star formation to low levels before it is shut off \citep[e.g., starvation;][]{Peng15}. We argue that the latter are better thought of as ``normal'' (uninterrupted) galaxy evolution, but accept the usage because it is so widespread.}

Quenching has the virtue of posing a precisely framed question: What agents truncate a galaxy's otherwise ``ordinary'' life process of continual growth via star formation? A drawback is that many candidates act on scales too small to simulate with current techniques \citep[e.g.,][]{Hopkins14}, have many possible parameterizations leading to widely varying effects \citep[e.g.,][]{Knebe15,Elahi15}, and rely on phenomena that are typically difficult to observe.

Such issues and the results of our own analyses have led us to an alternative paradigm. Over the course of a number of papers---\citet{Dressler13, Oemler13b, Gladders13b, Abramson13, Abramson14a, Abramson15}---we have developed a framework in which galaxy star formation histories [SFHs, records of in situ stellar mass $(\Mstel)$ production] are not maintained or squelched, but {\it extended} or {\it compressed}, rising and falling smoothly over time. The cessation of star formation is thus an extension of the {\it slowing} of star formation, and so the natural conclusion of the galaxy lifecycle, not the untimely termination of an otherwise healthy existence.

Seen in this light, focus shifts from local quenching mechanisms to some more-fundamental driver of {\it diversity} in growth histories. Provided the term can be defined to encompass gradual declines in star formation activity, we do not dismiss quenching {\it per se}, but suggest it to be a manifestation of this deeper organizing principle (Section \ref{sec:ICs}).

Here, we develop this argument by exploring the simple mathematical description of galaxy evolution detailed in \citet[][hereafter G13]{Gladders13b}. Though it neither is nor is meant to be a substitute for physical explanations, we suggest that its success in reproducing suites of data it was not designed to match either recommend it as a viable {\it paradigmatic description} of galaxy evolution, or question the utility of a number of cornerstone observations in this field.

Below, Section \ref{sec:worldview} outlines the motivation and construction of the G13 model. Section \ref{sec:phenotype} compares it to key observations---$\langle\ssfr(z)\rangle$, $d\log\sfr/d\log\Mstel$, stellar mass functions, ``fast-'' and ``slow-track'' quenching, the {\it UVJ} diagram, downsizing---and demonstrates its success at matching most of them. Sections \ref{sec:genotype}--\ref{sec:newSpaces} dissect the model to see why it behaves the way it does, concluding that (1) its interpretation of the scatter in the $\sfr$--$\Mstel$ relation as {\it Hubble timescale} SFH diversity, and (2) the physical emphasis it places on SFH {\it widths} are critical. Section \ref{sec:discussion} presents these findings' broader implications, namely that (1) ``grow-and-quench'' is an unnecessarily narrow framework of galaxy evolution, (2) initial conditions may be highly predictive of a galaxy's SFH, or (3) different kinds of data than are currently available are needed to show otherwise. Section \ref{sec:summary} summarizes.

Readers interested in our core arguments can jump to Sections \ref{sec:genotype}, \ref{sec:newSpaces}, or \ref{sec:discussion}. They might also see Appendices \ref{sec:AA}, \ref{sec:AB}, and \ref{sec:AC}, which flesh out key ideas presented in those Sections. Readers less interested in historical context can skip Section \ref{sec:motivation}.

We begin by revisiting the inception of the G13 model and its achievements to date.

\begin{figure*}[t!]
\centering
\includegraphics[width = 0.67\linewidth, trim = 0.5cm 0.5cm 0.5cm 0cm]{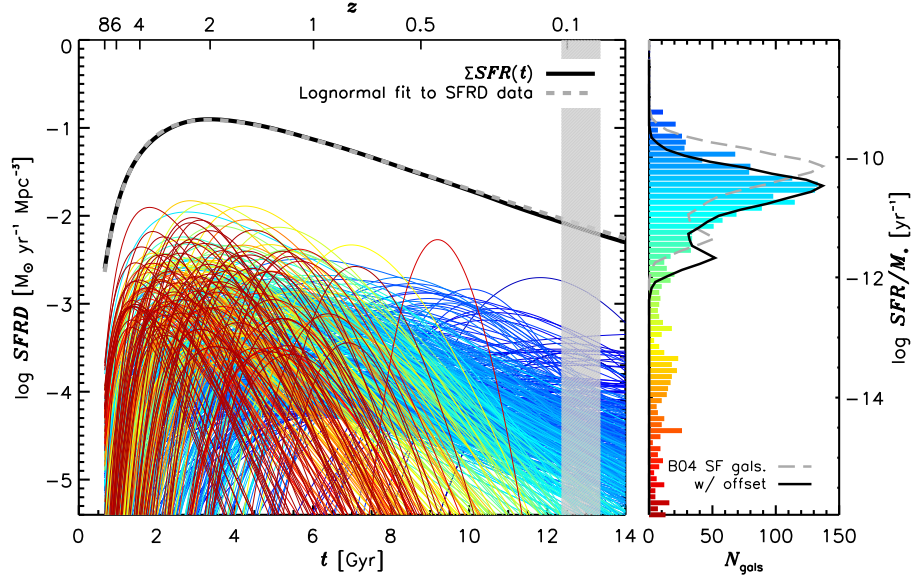}
\caption{G13 generates one lognormal SFH (colored lines at {\it left}; scaled by $10\times$ for visibility) for each galaxy in a volume-complete sample so as to reproduce their measured $\ssfr$s [{\it right}; SDSS $\ssfr$ distribution overlaid \citep{Brinchmann04}], and the cosmic SFRD$(z\leq8)$ (solid black and dashed grey curves, {\it left}). Input galaxies span redshifts shown by the shaded vertical band. We assume (1) the functional form of the SFHs, and (2) that non-starforming galaxies can have any $\sfr$ below some detection threshold ($0.05\, \Msun\, {\rm yr^{-1}}$; Figure \ref{fig:todaySFMS}, Equation \ref{eqn:sfrCons}; G13). Parameterizing the SFHs allows them to be projected to arbitrary lookback times. Any form can be used, but some are better than others (Section \ref{sec:contentFree}, Appendix \ref{sec:AA}; G13). The choice of the lognormal was motivated mainly by the shape of the cosmic SFRD (Figure \ref{fig:fit}). There is an 0.3 dex sSFR calibration offset between the B04 SDSS measurements and those used to constrain the G13 model. This is within systematic uncertainties (Figure \ref{fig:meanSSFR}) and is applied for visualization purposes in relevant comparisons (Figures \ref{fig:SFMStrends}, \ref{fig:delSFMS}).}
\label{fig:schema}
\end{figure*}


\section{Towards an Explicitly Quenching-Free Worldview}
\label{sec:worldview}

Of the papers mentioned in Section \ref{sec:intro}, \citet[][hereafter O13]{Oemler13b} and G13 are most germane to the present discussion. These works summarize the motivation and construction of the G13 model,  \resp.

\subsection{Motivation: A Diversity of Smooth {\rm SFH}s}
\label{sec:motivation}

O13 reached two conclusions based on analyses of specific star formation rate ($\ssfr\equiv\sfr/\Mstel$) distributions at $z\lesssim1$.

The first was that canonical SFH parameterizations---$\tau$ and delayed-$\tau$ models \citep{Tinsley72, Gavazzi02}---could not explain the global decline in $\ssfr$s observed over the above interval: they cannot generate both the tail of high $\ssfr$s at $z=1$ and the low values seen today. This problem---central to understanding how G13 bypasses explicit quenching (Section \ref{sec:sigms}, Appendix \ref{sec:AA})---is not small: $\sim25\%$ of $z\approx0$ galaxies with $\Mstel \gtrsim 4\times10^{10}\,\Msun$---a fair fraction of the Universe's stellar mass---cannot have evolved via the above prescriptions \citep[see also][]{Dressler16}. 

The second conclusion was that large discontinuities---e.g., starbursts---cannot be invoked to remedy this discrepancy. Such events (briefly) modulate the $\ssfr$s of individual galaxies, but they cannot drive the evolution of that quantity for whole populations over many Gyr (as is necessary): if some objects are in high states, others are in low ones, largely nulling their global effect (O13 Figure 4).

O13 thus demonstrated that many galaxies must have had SFHs that rose and fell rapidly at relatively late times (i.e., faster than the $\sfr$--$\Mstel$ relation at $z\lesssim1$) but were not starbursts.\footnote{The existence of such SFHs implies that starforming galaxies of the same $\Mstel$ do not grow up together/evolve homogeneously along the $\sfr$--$\Mstel$ relation. We argue that the latter scenario is both overly constraining and inaccurate (Sections \ref{sec:results}, \ref{sec:discussion}; \citealt{Dressler16}).} Taken with the results of, e.g., \citet{Rodighiero11} and our other investigations \citep{Dressler13,Abramson13}, these findings strongly suggested that abrupt discontinuities---upwards in the form of starbursts, or downwards in the form of rapid quenching---were not the primary shapers of global star formation over the past $\gtrsim7.5$ Gyr.

Instead, the smooth, long-timescale growth modes must encode important physics. Since O13 showed that their previous $\tau$/delayed-$\tau$ descriptions poorly captured this physics, a new SFH form was needed. 

G13 provided such a form and universalized the above conclusions into a general description of galaxy evolution.

\subsection{Construction: A [Log-]Normal ``Model'' of \\ Galaxy Evolution}
\label{sec:construction}

G13 recognized that the Universe's SFH---the evolution of the cosmic SFR density \citep[SFRD; e.g.,][]{MadauDickinson14}---is well described by a lognormal in time (Figure \ref{fig:fit}). That is:
\beq
	{\rm SFRD}(t) = \frac{A_{\rm Uni}}{\sqrt{2\pi\tau_{\rm Uni}^{2}}}\frac{\exp\left[-\frac{(\ln t - T_{0,\rm Uni})^{2}}{2\tau_{\rm Uni}^{2}}\right]}{t},
\label{eq:SFRDshape}
\eeq
where $T_{0}$, $\tau$ are the SFH's half-mass-time and width [in units of $\ln(\rm time)$], and $A$ is a scaling factor. G13 then asked a simple question: What if galaxy SFHs shared this functional form?

To explore this question, G13 took 2094 local galaxies with $\log\Mstel\geq10$ from the Sloan Digital Sky Survey \citep[SDSS; ][]{York00} and Padova Millennium Galaxy and Group Catalog \citep[PM2GC; ][]{Calvi11} and assigned them lognormal SFHs by fitting for the 2094 $(T_{0},\tau)$ pairs\footnote{For each galaxy, $A$ is set such that $\int_{t_{0}}^{t^{\rm obs}}\sfr(t)dt = \Mstel^{\rm obs}/f$, where $f$ is a $\sfr\mapsto\dot\Mstel$ conversion factor (0.7 for a Salpeter IMF).} that best reproduced (1) each galaxy's observed $(\Mstel,\sfr)$ while (2) ensuring that the ensemble of SFHs summed to the SFRD as far back as the data allowed ($z\leq8$; see Figure \ref{fig:schema}).

G13 and Appendix \ref{sec:AA} present technical details, but the above captures the model's essence: a continuum of smooth SFHs rising and falling naturally over loosely constrained timescales and peaking at loosely constrained times. No physical prescriptions govern these behaviors, only the Universe's observed SFH and the end-states of a set of real objects. 

In this sense, the G13 ``model'' is not a model at all, but a realization of possible lognormal SFHs for a sample of galaxies. As such, all physics must be inferred from the resulting $(T_{0},\tau)$ distribution, or read into the form of the SFH. Sections \ref{sec:genotype}, \ref{sec:newSpaces}, and \ref{sec:discussion} revisit this point.

Note: Given its $\log\Mstel(z\approx0)\geq10$ mass limit, G13 does not describe the SFHs of local dwarfs (though neither does it contain any). Unlike giant systems \citep{Rodighiero11}, these are known to grow via stochastic bursts \citep{Weisz14} which a single lognormal cannot capture. Hence, the following discussion is likely not applicable to them. Our principal interest lies in understanding what shapes the longer-term growth modes of galaxies that have consistently dominated the Universe's mass and $\sfr$ budget (Section \ref{sec:motivation}).

\subsection{Previous Descriptive Successes}
\label{sec:pastAchievements}

In G13, we showed that the model naturally reproduced the population of galaxies O13 (and subsequently \citealt{Dressler16}) found to have both late-peaking and narrow SFHs, something impossible to do using $\tau$ or delayed-$\tau$ SFH parameterizations (Figures \ref{fig:schema}, \ref{fig:todaySFMS}; O13). This encouraged us that the G13 model represented not only an improved description, but perhaps a physically meaningful worldview. 

More encouragement has come as the model continues to reproduce a range of observations it was neither constrained by nor intended to describe. We explore most of these results below, but review here what we have demonstrated to date:

\bitem
 	\item Constrained only by the cosmic SFRD and $z\approx0$ input galaxies, G13 (Figure 7) showed that the model faithfully reproduced $\ssfr$ distributions at four epochs at $z\lesssim1$. As O13 did for $\tau$/delayed-$\tau$ models, G13 showed that Gaussian SFHs failed this test badly.
	\item Turning the $z\lesssim1$ $\ssfr$ distributions into modeling constraints and refitting for each galaxy's $(T_{0},\tau)$, G13 (Figure 13) showed that the model reproduced the $z\sim2$ $\sfr$--$\Mstel$ relation (the ``Star Formation Main Sequence,'' or ``\MS'') from \citet[][]{Daddi07}.\footnote{The agreement with $z\lesssim1$ $\ssfr$ distributions achieved by the initial model suggests that the $z\sim2$ \MS\ could have been reproduced without turning these into constraints. We have not performed this test, however.} 
	\item Similarly constrained at $z\lesssim1$, \citet{Abramson15} showed that the model reproduced the evolution of the stellar mass functions of starforming and non-starforming galaxies at $z\lesssim2.5$. This was despite the fact that no mass-sensitive constraints were ever imposed except via the intrinsic mass distribution of the $z\approx0$ input data.
\eitem

These achievements by no means make the G13 model uniquely successful: others are equally capable \citep[e.g.,][]{Sparre14}. {\it However}, as G13 entails no explicit treatment of any physical process, we are spurred to investigate what makes this simple mathematical description an accurate expression of complex, poorly understood underlying physics.

\subsection{New Challenges}
\label{sec:newChallenges}

Here, we test the G13 model in a variety of new contexts to better ascertain its standing as a serious paradigm of galaxy evolution. We contend that its successes demonstrate its viability. But, at a minimum, they highlight issues at the descriptive level in this field that deserve real attention, or call for a reevaluation of some of its cornerstone observations.

\subsection{A Note to the Reader}
\label{sec:glossary}

Some vocabulary used below is unconventional in astronomy. Two definitions are in order:

\bitem
	\item[] {\bf Longitudinal data} -- repeated observations of individual objects---and only those objects---over a period of time. G13 describes such data (Section \ref{sec:construction}). Unfortunately, given the timescales of galaxy evolution, astronomers have no access to this information. Instead, we approximate it by time-ordering {\it cross-sectional} data.

	\item[] {\bf Cross-sectional data} -- observations of a sample of objects at a single epoch. We have no recourse but to assemble series of these data for different objects at different times to constrain galaxy evolution, so we must also use them to evaluate any and all models (Section \ref{sec:phenotype}). Sections \ref{sec:newSpaces} and Appendix \ref{sec:AB} discuss implications of this fact.
\eitem


\section{Results}
\label{sec:results}

We examine the G13 model from three different angles: 
\benum
	\item Section \ref{sec:phenotype} deals with its ability to reproduce important metrics in galaxy evolution neither used in its construction nor discussed in our previous work. 
	\item Section \ref{sec:genotype} explores how its core premises/traits guide these behaviors. 
	\item Section \ref{sec:newSpaces} projects the model into new spaces and uses the results to build a physical interpretation of both G13's behavior and its axiomatic underpinnings. It also presents some concrete predictions that may be testable with (future) observations to further assess the viability of the paradigm.
\eenum

We split the analysis in this way to clarify which aspects of the results reflect ``superficial'' details of the model's current realization and which are expressions of its underlying structure---its ``DNA.'' We assume $(H_{0},\Omega_{m},\Omega_{\Lambda}) = (73\, \kms\,\Mpc^{-1}, 0.27, 0.73)$ and a \citet{Salpeter55} initial mass function (IMF). 

\subsection{Comparing G13 to Data}
\label{sec:phenotype}

\subsubsection{The $\langle\ssfr\rangle$ of Low-Mass Galaxies Since $z=7$} 
\label{sec:meanSSFR}

Large photometric surveys have provided a wealth of data on the mean $\ssfr$ of relatively low-mass galaxies---$9\lesssim\log\Mstel\lesssim10$---at $z\lesssim7$ \citep[e.g.,][]{Noeske07, Daddi07, Damen09, Reddy09, Stark13, Gonzalez14}. These objects reside on the ``flat part'' of the $\ssfr$--$\Mstel$ relation, where $d\log\ssfr/d\log\Mstel\simeq0$ \citep[or $d\log\sfr/d\log\Mstel\simeq1$; e.g.,][]{Whitaker14}. $\langle\ssfr\rangle$ is thus akin to the \MS\ zeropoint.

In this mass regime, the only G13 modeling constraint was the cosmic SFRD$(t)$; i.e., the evolution of the sum of SFRs of all galaxies. Additionally, our understanding of $\langle\ssfr(z)\rangle$ has evolved markedly since the construction of the model \citep[][]{Stark13,Gonzalez14,Kelson14}. For both reasons, these data provide a meaningful test of the accuracy of the G13 SFHs and therefore the validity of the approach. 

\begin{figure}[t!]
	\centering
	\includegraphics[width = \linewidth, trim = 0.5cm 0cm 0cm 0cm]{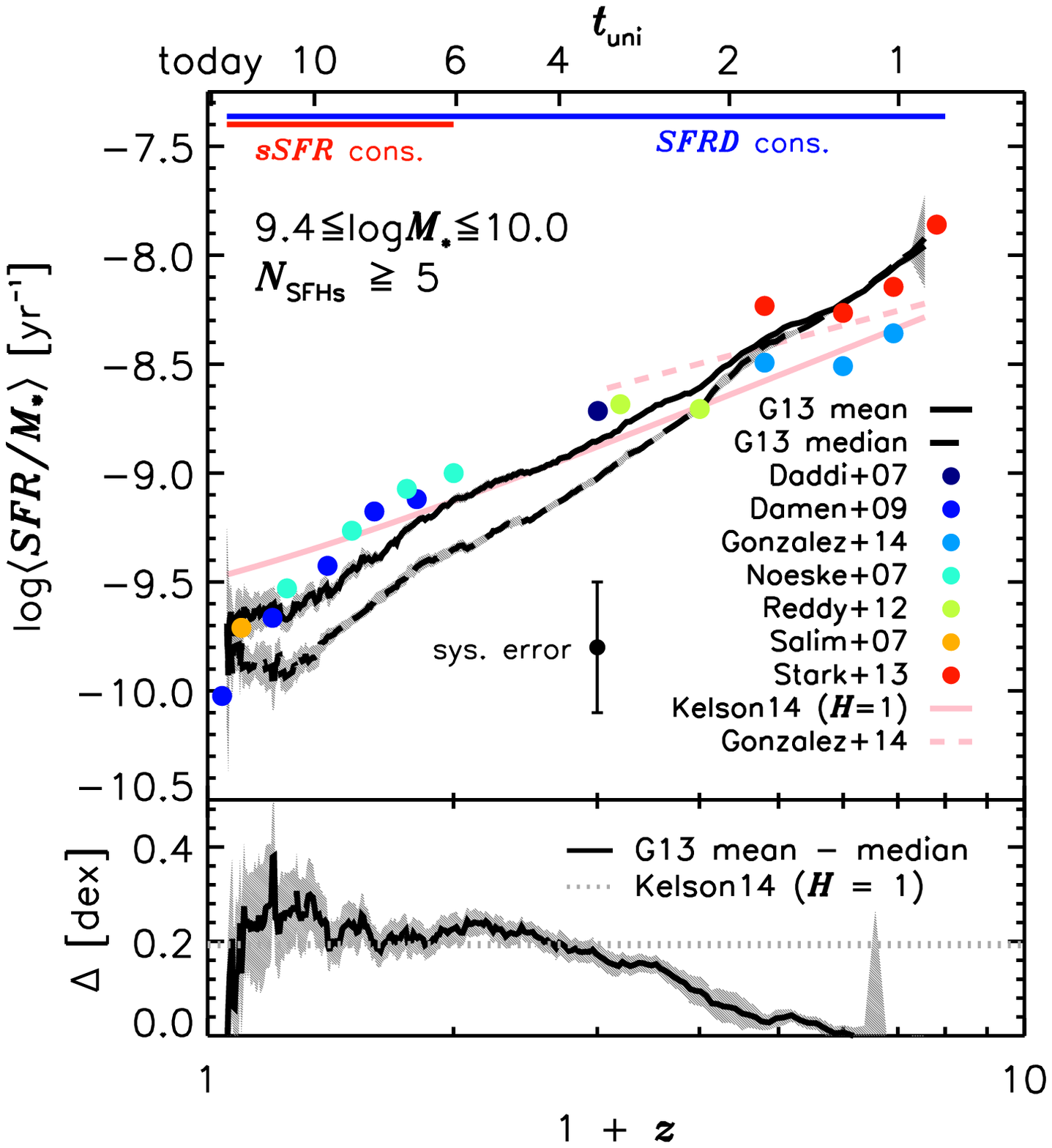}
\caption{{\it Top}: mean/median $\ssfr(9.4\leq\log\Mstel\leq10.0)$ from G13 (solid/dashed black lines), describing systems on ``the flat part'' of the \MS\ where $\ssfr$ is mass-independent. Model measurements span epochs where at least 5 SFHs support it. Data (colored points; estimates of the {\it mean}) and systematic uncertainties (black point) are from \citet{Gonzalez14} and \citet{Salim07}. Agreement is good everywhere, especially considering that the model is unconstrained at $z>1$ except by the cosmic SFRD (blue bar at top), and that data from \citet{Stark13} and \citet[][]{Gonzalez14} were published after the model was created. {\it Bottom}: the mean--median offset, a probe of the \MS\ scatter. To $z\sim2$, the model reproduces the $\sim0.2$ dex offset implied by K14's preferred stochastic evolutionary scenario (see text; his Figure 3{\it a}), showing that paradigms with opposite assumptions can produce the same (non-)evolution in this quantity.}
\label{fig:meanSSFR}
\end{figure}

Figure \ref{fig:meanSSFR} presents this test. Here, we show how the $\langle\ssfr\rangle$ of SFHs with $9.4\leq\log\Mstel(z)\leq10$ compares to data for galaxies in the same mass range taken from \citet{Gonzalez14} and \citet{Salim07}. Agreement is remarkably good at all $z\leq7$ (where at least 5 SFHs support the model measurement), especially considering systematic uncertainties (black point), and that the points from \citet{Stark13} and \citet{Gonzalez14} were published after the model was created. In fact, at the time of its construction, $\langle\ssfr\rangle$ was thought to flatten at $z>2$ \citep{Stark09}; only later, when  SED fits were adjusted to account for emission line fluxes \citep{Stark13}, did those points show the monotonic increase anticipated by G13. 

We neglect mergers here and in all following Sections. Results from \citet{Leitner12}, \citet{Behroozi13}, \citet{Abramson15}, and our own numerical tests suggest that this move does not affect the arguments below, but it is worth keeping in mind. For example, as presented (see Appendix \ref{sec:AA}), the SFHs may average over the histories of multiple pre-merged systems at high-$z$. Assuming such pieces have similar $\ssfr$s---as G13 did and is implied by the \MS\ slope (see above and Section \ref{sec:MSslope})---model galaxy counts would be impacted more than $\sfr$-related metrics (Section \ref{sec:SMFs}).

Figure \ref{fig:meanSSFR} is meant to be compared with Figure 3{\it a} of \citet[][hereafter K14]{Kelson14}. That model predicts a similar trend to G13 (pink line) but adopts a very different physical premise, positing that individual galaxies evolve via quasi-stochastic $\sfr$ discontinuities correlated on arbitrary timescales denoted by $H$. $H=1$ corresponds to {\it Hubble timescale} correlations and fits the data well (K14 Section 3). It implies a {\it median} $\ssfr(t)\approx(H+1)/t = 2/t$, with a mean $\sim0.2$ dex higher. That offset is a measure of the \MS\ scatter, $\sigms$.

Notably, Figure \ref{fig:meanSSFR}, {\it bottom}, shows that, to $z=2$---three Gyr prior to the earliest $\ssfr$ constraint---the G13 model {\it also} produces a $\sim0.2$ dex mean/median offset. This is intriguing because K14 and G13 are based on not only different, but {\it opposite} assumptions: G13 assumes that SFHs are smooth, parametrizable, and diversified quasi-deterministically; K14 assumes that they are discontinuous, unparametrizable, and diversified by stochastic changes in equilibrium conditions. Section \ref{sec:descriptions} revisits the potential ramifications of such comparably accurate yet physically distinct descriptions.

Moving to higher-$z$, the G13 mean and median $\ssfr$s converge, implying a decrease in $\sigms$. This result is inconsistent with recent analyses that show it to be roughly constant instead of shrinking \citep[e.g.,][]{Schreiber14,Speagle14}. Yet deciphering this discrepancy---and some others that will arise later---is non-trivial.

First, it may be superficial, not structural, at least over the timespan probed. The model offset is sensitive to the location of $\ssfr$ distributional constraints, and so could be changed by imposing additional ones at $z>1$.

\begin{figure}[t!]
	\centering
	\includegraphics[width = \linewidth, trim = 1.2cm 0.5cm 0.5cm 0cm]{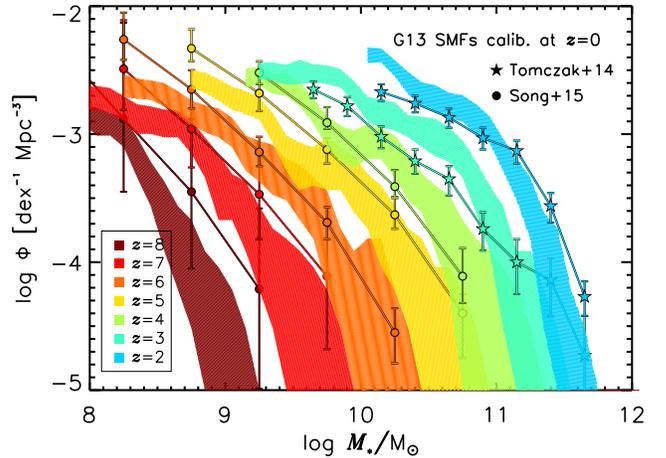}
	\caption{G13 stellar mass functions (colored bands; 90\% credibility interval) at $z\leq8$ compared to data from \citet{Song15} and \citet{Tomczak14} (points). The model is calibrated only to $z\approx0$ data \citep{Moustakas13} over the range $\log\Mstel\in[10,10.5]$ (Figure \ref{fig:SMFincompl}), and constrained only by the cosmic SFRD over all epochs plotted.}
\label{fig:SMFs}
\end{figure}

Second, it may be accurate: recall that the data reflect cross-sectional samples (different objects at different times; Section \ref{sec:glossary}), which, at $z>0$, will almost certainly contain objects that do not represent progenitors of the G13 input sample (Sections \ref{sec:construction}) and so are not described by the model. Hence, it may be that, when true progenitors are isolated {\it a priori}, their $\ssfr$s do converge at early times, with non-ancestral galaxies ``padding-out'' the scatter in the data (Figure \ref{fig:delSFMS}).

That said, at some point a decrease in $\sigms$ (but not necessarily that shown in Figure \ref{fig:meanSSFR}) must become structural: At $t\ll\langle\tau\rangle\sim1.6$ Gyr---the mean G13 characteristic timescale for the ensemble of SFHs---G13 ``galaxies'' share approximately the same exponential $\sfr(t)$, leading to a unique SFH for each $\Mstel(t)$ (which becomes a scale factor). So, if $\sigms$ is ultimately shown {\it not} to decrease at those times ($z\gtrsim3$--4)---or if the addition of $\ssfr$ constraints at $z>2$ cannot remedy any real discrepancies there---we will learn something important about the limitations of smooth (lognormal) SFHs.

\subsubsection{The Evolution of the Stellar Mass Function Since $z=8$}
\label{sec:SMFs}

Besides the evolution of the mean $\ssfr$ of (low-mass) galaxies, another critical observation to reproduce is the evolution of the stellar mass function (SMF). \citet{Abramson15} demonstrated the success of the G13 model in this context at relatively low redshifts ($z\lesssim2.5$). We now extend this test to earlier epochs.

\begin{figure}[t!]
	\centering
	\includegraphics[width = \linewidth, trim = 0.5cm 0.5cm 1cm 0cm]{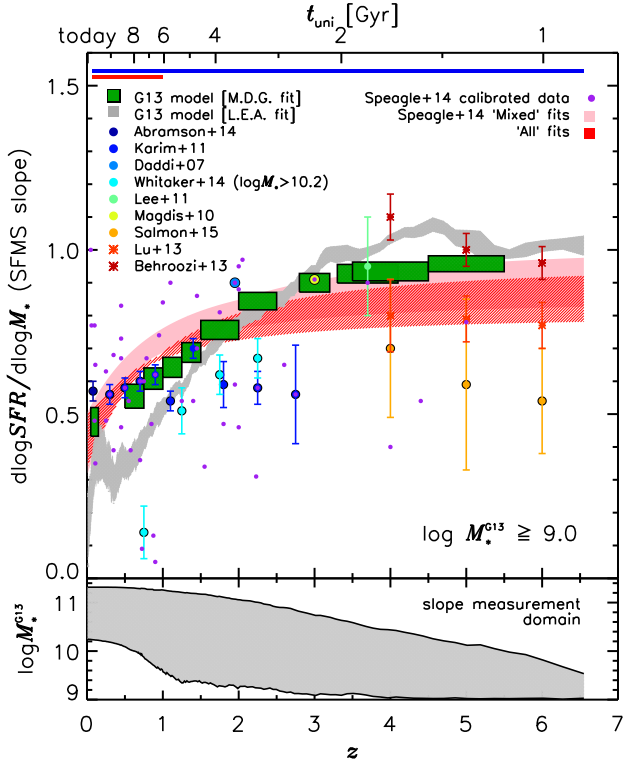}
	\caption{{\it Top}: G13 predictions for the evolution of the slope of $\sfr(\Mstel)$ (green boxes, grey shaded band) compared to measurements from the literature (colored circles) and semi-analytic models (stars). Systematic uncertainties are large, but a general trend toward more-negative values at later times is apparent in both the data and G13 [see \citealt{Speagle14}'s extensive analysis; pink band]. {\it Bottom}: the $95\%$ G13 $\Mstel(t)$ range considered by L.E.A., who continuously fit the SFHs as they evolved. M.D.G. bootstrapped them in redshift windows similar to those in the data and used a different $\sigma$-clipping scheme. The fact that there is a sensible model trend at all is unexpected.}
\label{fig:SFMSslope}
\end{figure}

Figure \ref{fig:SMFs} plots the G13 prediction for the evolution of the total SMF at $2<z\lesssim8$, as far back as data are available. The only calibration applied to the model is an absolute $y$-offset to match the $z\approx0$ SMF of \citet{Moustakas13} at $10\leq\log\Mstel\leq10.5$ (10 Gyr away from the earliest epoch plotted), and the removal of some over-represented high-mass galaxies in the G13 input sample (Appendix \ref{sec:AD}, Figure \ref{fig:SMFincompl}).

Considering that no mass functions were used to constrain the model (except through the intrinsic mass distribution of the $z\approx0$ input sample), agreement is remarkably good over more than a decade in $\Mstel$, two decades in space density, and lookback times approaching 13 Gyr. This finding substantially extends those of \citet[][]{Abramson15} regarding the accuracy of G13 SFHs at the mass-bin level.

A systematic offset may exist at $z\sim 8$, but we caution against over-interpreting it: At all epochs shown, the model is constrained only by the sum of all SFHs. Hence, there is no obvious reason to expect the shape or normalization of the SMFs to look anything like the data (something we have confirmed through numerical tests). Additionally, we have not verified that the highest-$z$ measurements---convolved with the appropriate $\sfr$--$\Mstel$ relation---are in fact consistent with the SFRD used to derive the model, which only runs to $z=8$. Finally, assuming the data are reliable and consistent, and that the offset is meaningful, this tension is again probably superficial; it 	should be eased by introducing further constraints at $z>1$.\footnote{The volume probed by, e.g., \citet{Tomczak14} is $\sim10\times$ that of PM2GC (used by G13). Our results are thus subject to shot-noise in terms of the area of $(T_{0},\tau)$ space the data {\it must} span: The input sample may contain too few massive (red) galaxies to guarantee very-early-peaking output SFHs. Rerunning the model on a larger volume might therefore also remedy this issue, but is beyond the scope of this work.} Indeed, a comparison of G13 Figures 5 and 9 shows that adding $\ssfr$ constraints tends to push (subsets of) SFHs towards earlier peak-times, precisely as the data suggest. 

For all of these reasons, we view this result as an unanticipated success of the G13 model.

At low-masses, the G13 SMFs are truncated where they turn over (Appendix \ref{sec:AD}, Figure \ref{fig:SMFincompl}). Because of the longitudinal nature of our sample (it tracks the same group of objects through time; Section \ref{sec:glossary}), this completeness threshold evolves {\it downward} with increasing redshift. At high-masses, a cut-off is imposed by the location of the most-massive model ``galaxy'' at any epoch. Increasing the number of input galaxies to which SFHs are assigned will modulate both bounds, as will incorporating mergers, which remove high-mass and add low-mass systems. Figure \ref{fig:SMFincompl} provides more context for how these effects might enhance or degrade agreement with data.

A final note: The evolution of the \MS---and therefore $\langle\ssfr(z)\rangle$---is linked to that of the cosmic SFRD via the galaxy stellar mass function \citep[at least for starforming systems; ][]{Abramson15}. Since the G13 model is designed to match the SFRD, and, as shown in Section \ref{sec:meanSSFR}, also captures the evolution of $\langle\ssfr(z)\rangle$ over most of the interval spanned in Figure \ref{fig:SMFs}, a concern might be that it is guaranteed to capture the evolution of the SMF. This is only true, however, over the mass range $9.4\lesssim\log\Mstel\lesssim10$ {\it at epochs where such galaxies dominate the} SFRD. There is nothing in the modeling procedure that guarantees matching the rest of the SMF at those epochs, or any of it at other epochs. That we accomplish the latter suggests that the model must correctly reproduce the evolution of the entire \MS. We explore this next.

\begin{figure*}[t!]
	\centering
	\includegraphics[width = 0.9\linewidth, trim = 1.5cm 0.2cm 1cm 0cm]{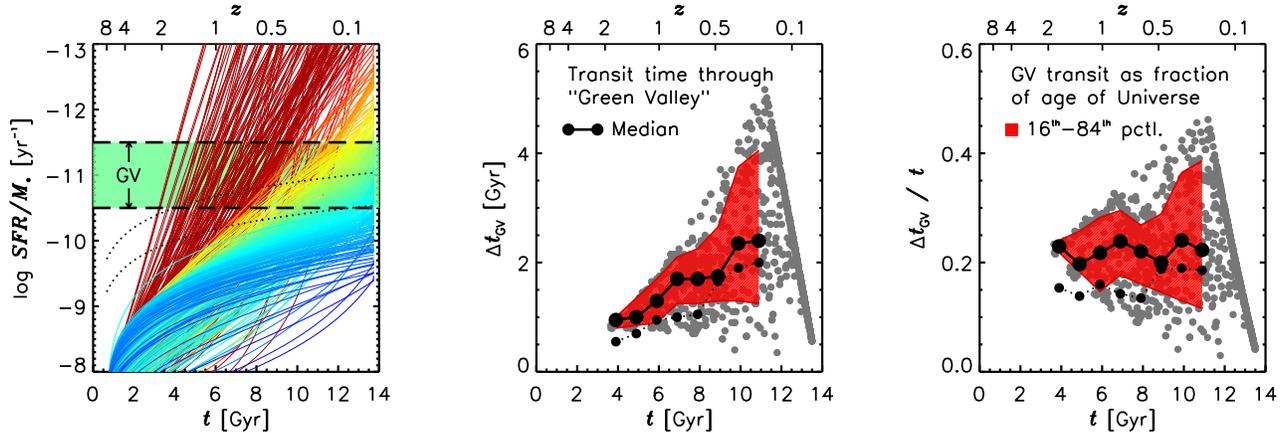}
	\caption{G13 naturally describes the shift from so-called ``fast-'' to ``slow-track'' quenching. {\it Left}: model $\ssfr$s as a function of time colored by present-day values [red = low, blue = high; $y$-axis oriented as in \citet{Barro13} and the color/mass diagram]. A fiducial transition zone between SF and quiescent galaxies---the ``Green Valley'' (GV)---is highlighted by the green shaded band with an alternative---evolving $\propto1/t$ with half the width in $\log\ssfr$---plotted as dotted lines. Slopes of SFHs crossing the GV are steep at high-$z$ but shallower today, implying quenching timescales were rapid in the past and are slower now. The {\it middle} panel plots GV transit time---$\Delta t_{\rm GV}$---against cosmic time to quantify this trend; small dots show medians using the $1/t$ GV definition. Interestingly, while G13 recovers fast- and slow-track quenching, once normalized by the age of the Universe at crossing ({\it right}), {\it the absolute trend disappears}: galaxies spend $\sim20\%$ of their life in the GV. This jibes with the scale-free interpretation of K14 and also supports the idea that non-starforming galaxies have compressed SFHs, but have {\it not} been subject to ``special'' physics as a class. (See also Figure 11 of \citealt{Vulcani15a}.)}
\label{fig:GVtransit}
\end{figure*}

\subsubsection{The Evolution of the {\rm \MS} Slope Since $z=6$}
\label{sec:MSslope}

Figure \ref{fig:meanSSFR} tracked galaxies with $\log\Mstel(z)<10$, where effectively all stellar mass is associated with star formation (e.g., in a disk) and $d\log\ssfr/d\log\Mstel\approx0$. These define the \MS's zeropoint. Now we examine the evolution of its slope.

To some extent, the \MS\ is affected by the presence of bulges in high-mass galaxies, which add $\Mstel$ at fixed $\sfr$. This effect would tend to push the \MS\ slope to more-negative values (\citealt{Abramson14a, Whitaker15}; Oemler et al., in preparation, Figure \ref{fig:BTstuff}, below; but cf.\ \citealt{Schreiber16}, who attribute the effect to reduced SF efficiency). As it is now believed that bulge growth could be a multi-phase process, with some galaxies forming them at late times \citep[e.g.,][]{Lang14,HuertasCompany15}, the slope of $\sfr(\Mstel)$ should monotonically evolve from $\sim1$ at high-$z$ to lower values today if fit by a single power-law.

Figure \ref{fig:SFMSslope} shows data (points), drawn mainly from Table 4 of \citet{Speagle14}. As substantiated by those authors' thorough meta-analysis (pink/red bands), the slope indeed evolves subtly from $d\log\sfr/d\log\Mstel\sim0.8$ at $z\approx4$ to $\sim0.4$ today.

As for G13's predictions, independent fits based on (1) continuous measurements of the SFHs (grey band; performed by L.E.A.) or (2) bootstrapping in redshift windows defined by the data (green boxes; performed by M.D.G.) show the slope to evolve similarly, dropping from near unity at $z\approx4$ also to $\sim 0.4$ today.\footnote{Both approaches limited fits to SFHs with $\log\Mstel(z)\geq9$---a lower bound encompassing almost all data sets. Discrepancies here illustrate the difficulty of measuring this quantity robustly.} Beyond $z\approx4$, the G13 model is broadly consistent with both the \citet{Speagle14} projections, and results from at least one semi-analytic model based on dark mater halo growth (\citealt{Behroozi13}; perhaps less so \citealt{Lu14a}; both plotted as colored stars). 

Clearly, as evident from comparing points at similar epochs, systematic uncertainties in the data are large at all $z$ due, e.g., to differences in mass ranges and definitions of ``starforming.'' These remain despite attempts to calibrate-out variations in measurement techniques (purple points; see Table 6 of \citealt{Speagle14}). Also, the (scant) extant measurements at $z>4$ \citep{Salmon15, Steinhardt14a} are generally lower than model predictions, so more and better information is needed to fully assess any findings here.

Nevertheless, a core point is robust: The G13 model neither contains physical prescriptions (for bulge growth or star formation efficiency) nor is bound by constraints that would inform its behavior in Figure \ref{fig:SFMSslope}. In fact, at $z>1$, it ``knows'' only about the cosmic SFRD. As the mass-insensitive integral of the SFRs of all galaxies, this is perhaps as far removed from the \MS\ slope---sensitive to mass differentials between individual systems---as a constraint can be. Hence, as above, that G13 qualitatively reflects the data---and indeed might be quantitatively consistent with state-of-the-art analyses and semi-analytic models---either speaks to its validity, or comments on the physical information content of the metrics examined. Sections \ref{sec:newSpaces} and \ref{sec:contentFree} revisit this point.

\subsubsection{The Congruity of Fast- and Slow-Track Quenching}
\label{sec:GV}

So far, we have dealt with starforming galaxies. These dominate the stellar mass density of the Universe at $z\gtrsim1$ \citep[][]{Muzzin13,Tomczak14}, so limiting the discussion to them is not too biasing. However, since we have set the G13 model in opposition to those specifically designed to create {\it non}-starforming galaxies, it behooves us to discuss quenching in this framework. We do so now and in the next Section.

Recently, a picture has emerged in which galaxies transitioned from a starforming to non-starforming state more rapidly in the past than they do today (on average). This dichotomy between ``fast-track'' and ``slow-track'' quenching \citep{Barro13, Schawinski14} has been used to suggest an interesting change in the mechanisms inducing this transformation over cosmic time. The quenching timescale may therefore be a critical datum in galaxy evolution.

The G13 model suggests otherwise, however. Figure \ref{fig:GVtransit}, {\it left}, shows tracks of $\ssfr(t)$ for a representative subset of SFHs. As in \citet{Barro13}, the $y$-axis is oriented to correspond with the color-mass diagrams in which the active/passive transition is often discussed (red/low-$\ssfr$ {\it up}). The green horizontal swath and dotted tracks in $\ssfr$ approximate two definitions of the transition regime (see below); i.e., the ``Green Valley'' (GV). The middle panel shows the time each SFH spends in the GV; i.e., the quenching timescale, $\Delta t_{\rm GV}$.

Two things are clear. First, the G13-derived quenching timescale is indeed an increasing function of time, with transformations being faster in the past than they are today (middle panel). Crossing times depend quantitatively on the definition of the GV, but---as shown by choosing one that evolves $\propto 1/t$ (small dots)---the trend will not (for reasonable choices). Thus, G13 naturally reproduces the fast-track/slow-track quenching transition. However, this {\it need not correspond to a change in quenching physics} because the model contains no explicit quenching mechanism.

Instead, Figure \ref{fig:GVtransit}, {\it left}, makes it clear that the trend is purely a function of the slope of the SFHs. Upon reflection, this result is inevitable: To reach their final mass early---i.e., to become a passive system at high-$z$---galaxies {\it must grow rapidly}. Hence, their SFHs {\it must} rise and fall steeply, and $\Delta t_{\rm GV}$ must be short. Conversely, a galaxy crossing the GV today is likely to have had a much more extended SFH, and therefore a much more slowly evolving $\ssfr(t)$. Thus, it will naturally linger in the GV with larger $\Delta t_{\rm GV}$ relative to systems that quenched earlier \citep[see also][]{Pacifici13}. 

This phenomenon reflects a deeper point: different parts of $(T_{0},\tau)$ space support the SFRD at different times (Appendix \ref{sec:AC}). Hence, the SFHs corresponding to red or blue galaxies at {\it fixed} $(\Mstel,t)$ can derive from disjoint loci in G13's underlying unimodal $(T_{0},\tau)$ distribution: the former hail from regions dominating the past SFRD, the latter from regions dominating the present (Figures \ref{fig:TtauMass}, \ref{fig:bimodality}). Thus, continuous descriptions can manifest quenched/starforming bimodalities.

This interpretation---that the transition from fast- to slow-track quenching does not represent a change in mechanism, but the natural shift in the GV population from galaxies that are rapidly growing to those that are slowly growing---is driven home by Figure \ref{fig:GVtransit}, {\it right}. There, we replot $\Delta t_{\rm GV}$, but as a fraction of the age of the Universe mid-way through the crossing ($\Delta t_{\rm GV}/t$). Seen in this light, there is no evolution in the crossing time at all. Scatter between SFHs is large, but the median stays at roughly $20\%$ of the age of the Universe (effectively the age of a galaxy) forever. (\citealt{Zolotov15} find a similar 35\% using hydrodynamic simulations.) As such, quenching appears to be a scale-free phenomenon qualitatively consistent with the picture of K14 \citep[and perhaps][]{Stringer14}. If so, quenching timescales may yield little insight into evolutionary processes.

\citet{Barro13} tied fast- vs.\ slow-track quenching to galaxy sizes, with compact galaxies evolving via the fast route. As we show in Sections \ref{sec:moreTau} and \ref{sec:BT}, the G13 model naturally establishes this connection while maintaining an underlying scale-free nature to quenching through $\tau$.

\begin{figure*}[b!]
	\centering
	\includegraphics[width = 0.825\linewidth, trim = 0cm 0cm 0cm 0cm]{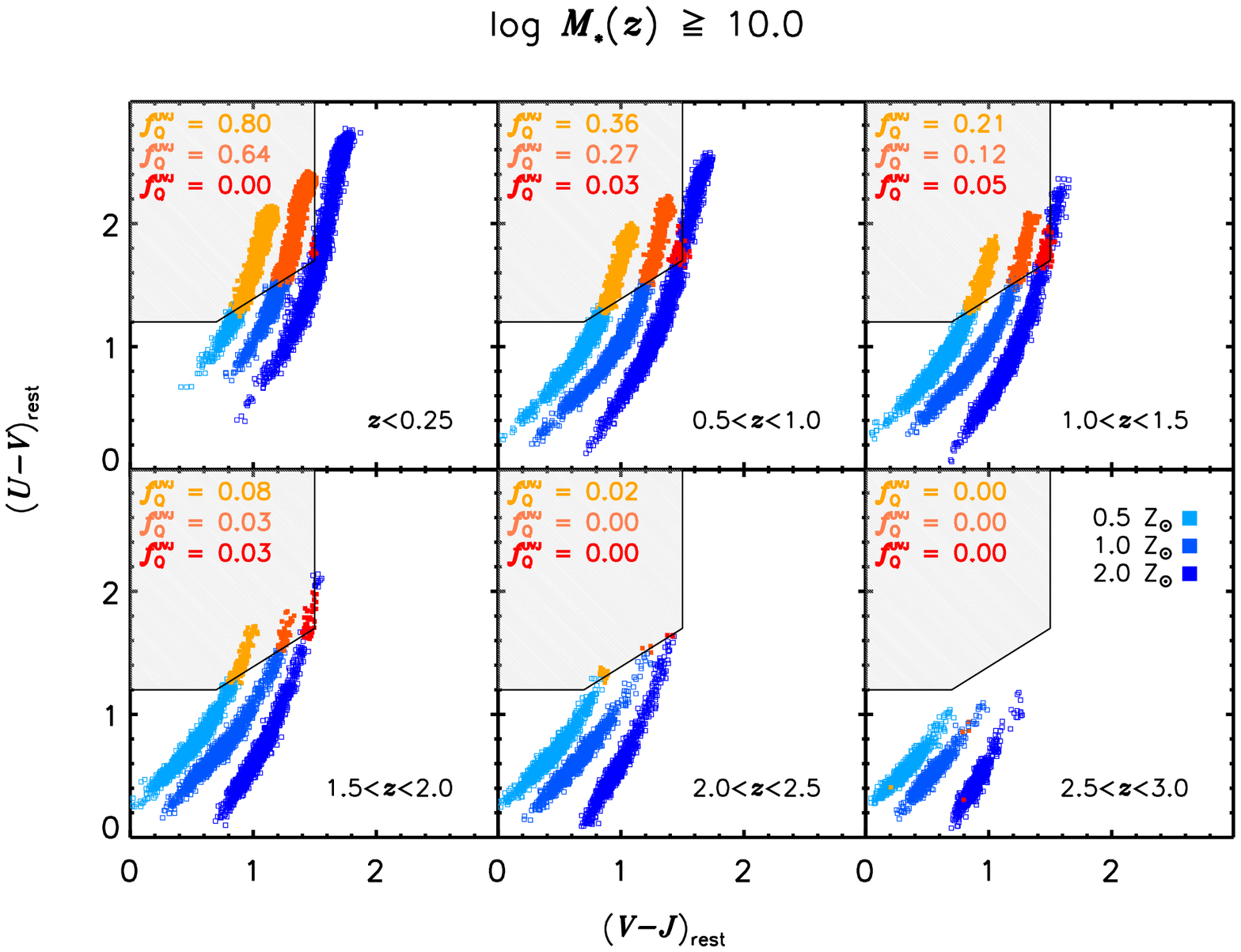}
	\caption{Synthetic color-color diagrams derived from the G13 SFHs and the models of \citet[][]{BC03} shown as in \citet{Tomczak14}. Blue/orange points show SFHs that would be classified as starforming/quiescent, \resp, using the {\it UVJ} criterion of that work. Each SFH was run assuming a fixed metallicity of $Z = \{0.5,1.0,2.0\}\times Z_{\odot}$ (light, medium, and dark shades, \resp). Only SFHs with $\log\Mstel(t)\geq10$ are shown to correspond with the mass range probed by \citet{Abramson15}. As printed in the top-left of each panel, at face value, the model underproduces quiescent systems at $z\gtrsim1.5$; integrating the \citet{Tomczak14} mass functions over this mass range yields $f_{\rm Q}^{\rm obs}\sim24\%$--28\% at these epochs. Yet, the model matches data if an \MS-relative $\ssfr$ cut is used \citep{Abramson15}, implying that this shortcoming could be ameliorated by introducing additional $\ssfr$ constraints.}
\label{fig:UVJ}
\end{figure*}

\begin{figure*}[b!]
	\centering
	\includegraphics[width = 0.9\linewidth, trim = 2cm 0cm 2cm 0cm]{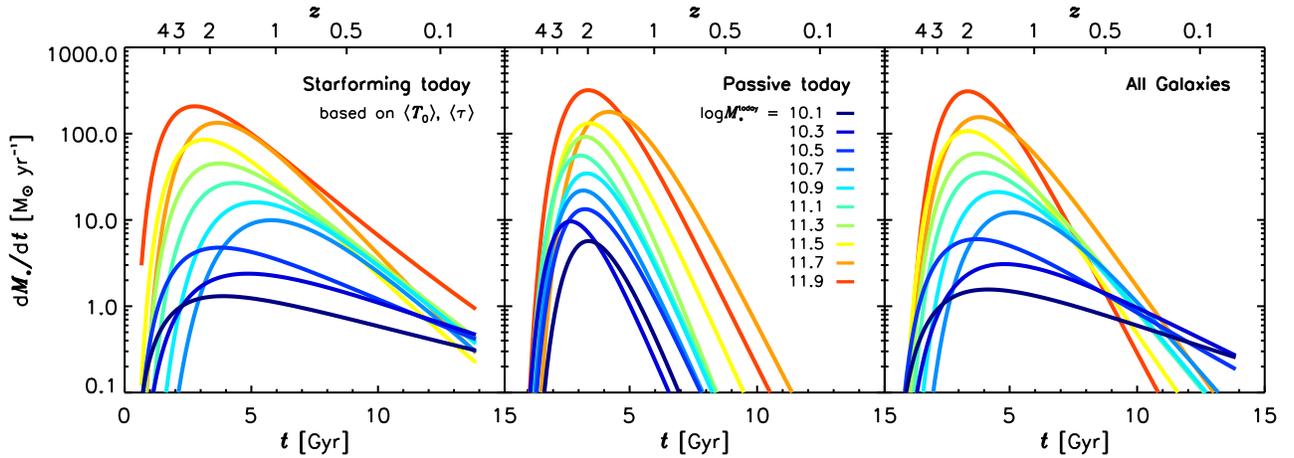}
\caption{Lognormal SFHs derived from the median $(T_{0},\tau)$ parameters in bins of present-day $\Mstel$. Panels show results considering all galaxies ({\it right}), or only those currently classified as starforming ({\it left}) or non-starforming ({\it middle}; cut at $\ssfr = 10^{-11}\, {\rm yr^{-1}}$). A ``downsizing'' trend is clear in the starforming population---driving that seen for all galaxies---with today's lower-mass systems forming later/with more extended SFHs than higher-mass systems. That trend is not visible in present-day passive G13 galaxies because these have effectively no $\ssfr$ constraint at late times. Hence, the SFHs are guided mainly by the cosmic SFRD, and so follow its shape closely regardless of final mass (note the consistent peak at $z\approx2$). In reality, some of these systems would have formed earlier than this. Such behavior leads to the under-abundance of objects in the passive region of the $z\gtrsim1.5$ {\it UVJ} diagram (Figure \ref{fig:UVJ}).}
\label{fig:medTtau}
\end{figure*}

\subsubsection{The {\rm UVJ} Diagram and Color Selection}
\label{sec:UVJ}

With the advent of large, high-redshift surveys, the division between starforming and non-starforming/quiescent/passive galaxies must be made according to photometric and not spectroscopic criteria. A popular distinction is based on the distribution of galaxies in $U-V$/$V-J$ color-color space---the ``{\it UVJ} diagram'' \citep[e.g.,][]{Williams09}. Hence, another useful test of the G13 model is to examine its predictions for the evolution of galaxies in this plane.

Figure \ref{fig:UVJ} shows the results of running the stellar population synthesis (SPS) code of \citet[][]{BC03} to extract synthetic colors for each SFH in different redshift intervals assuming three different metallicities---$Z\in\{0.5,1.0,2.0\}\times Z_{\odot}$. We limit the plotting to galaxies with $\log\Mstel(z)\geq10$ at any epoch to remain consistent with \citet{Abramson15}.

Taking the results at face value, the G13 model fares worse in this context than in the others explored so far. While, accounting for metallicity spreads, our results appear roughly consistent with, e.g., the \citet[][]{Tomczak14} data at $z\lesssim1.5$ ($0.28 \lesssim f_{Q} \lesssim0.47$), the model underproduces photometrically quiescent systems at higher-$z$, returning percentages in the single-digits as compared to the 24\%--28\% from that work. As such, had we split the galaxy population using this technique as opposed to an evolving \MS-relative $\ssfr$ cut, we would not have correctly reproduced the evolution of the quiescent mass function shown in \citet{Abramson15}.

However, because the model {\it does} reproduce the correct SMF evolution using an evolving $\ssfr$ cut, the lack of {\it red} objects would seem to reflect a mismatch in the mean shapes of SFHs leading to color-defined passive galaxies at those epochs, not a real deficit of ``quenched'' systems. That is, the comparatively low-$\ssfr$ systems we identified were ``quenching,'' just not rapidly enough to produce colors photometrically identifying them as such at early epochs. 

Thus, Figure \ref{fig:UVJ} may again reveal a superficial issue related to G13's data constraints: if a $z>1$ $\ssfr$ constraint was introduced---forcing some SFHs to lower $(T_{0},\tau)$ and thus steeper declines---a model with identical assumptions may well produce enough systems reaching low enough $\ssfr$s at early enough times to match the photometry.

The next Section elaborates on this statement to get at {\it why} the model deviates from the data in this context.

\subsubsection{Average Galaxy {\rm SFH}s and Downsizing}
\label{sec:downsizing}

First insights come from Figure \ref{fig:medTtau}, showing the average SFHs of galaxies split by present-day $\Mstel$ and $\sfr$.

When examining either all SFHs ({\it right}), or the subset that lead to ``starforming'' systems today ({\it left}; $\log\ssfr\geq-11$), a clear trend emerges for more-massive systems (redder curves) to grow/peak earlier than less massive systems (bluer curves). This model trend corresponds with the now well-established observational paradigm of galaxy ``downsizing'' \citep[e.g.,][]{Cowie96,Neistein06,Pacifici13}, and is directly related to the apparent slow-down in Green Valley crossing times discussed in Section \ref{sec:GV}.

The middle panel, however, reveals no such trend for G13 SFHs leading to today's non-starforming/passive galaxies [$\ssfr(z=0) < 10^{-11}\, {\rm yr^{-1}}$]. These histories are basically identical when scaled by their present-day $\Mstel$.

The reason the model produces this uniformity in passive galaxy SFHs has to do with the amount of $\ssfr$ ``real estate'' they can occupy compared to their starforming counterparts. As Figure \ref{fig:schema} shows, starforming galaxies today, by definition, must land on a relatively narrow $\ssfr$ target ($\sim 1$ dex) defined by the $z\approx0$ \MS. Passive galaxies, on the other hand, can occupy an effectively infinite space below this locus. Thus, they have essentially no late-time $\ssfr$ constraint. As the only other constraint in the model is the cosmic SFRD, the SFHs for these systems naturally tend towards that history, rising quickly to peak at $z\sim2$ and then falling as seen in Figure \ref{fig:medTtau}. Since colors cannot discriminate between objects that last formed stars, e.g., 3 vs.\ 10 Gyr ago---but are very good at identifying those that formed them within the past 0.1--1 Gyr---the ability for these SFHs to take arbitrarily low $z\approx0\, \ssfr$s need not translate to red colors at $z\gtrsim2$, allowing the corresponding area in {\it UVJ}-space to remain vacant.

The upshot is that, absent additional $\ssfr$ constraints, it is difficult for the G13 approach to produce large numbers of SFHs reaching low $\ssfr$s before the cosmic SFRD peaks (Figure \ref{fig:schema}). This leads to the dearth of red {\it UVJ} galaxies seen at $z>1.5$ (Figure \ref{fig:UVJ}) and probably contributes to any real underestimation of the $z\approx8$ SMF (Figure \ref{fig:SMFs}). As such, this behavior will again be modulated by adding $\ssfr$ constraints at higher-$z$, or by rerunning the fitting on a larger number of systems that presumably more-fully span $(T_{0},\tau)$ space.

A final note about downsizing: As is empirically well-established and shown in Figure \ref{fig:meanSSFR}, the \MS\ has fallen consistently over cosmic time. This means that galaxies doubled in mass (e.g.) faster in the past. As is also well-established and shown in Figure \ref{fig:SFMSslope}, the slope of the (high-mass end of the) \MS\ is sub-unity for most of cosmic time. This means that low-mass galaxies always double in mass faster than high-mass galaxies. These facts combine to ensure that galaxies growing-up at earlier times {\it rushed to states of comparative inefficiency} (``raced to the bottom'') more quickly than galaxies growing-up at later times. This {\it is} downsizing, and it is a generic prediction of any model leading to/incorporating $d\log\sfr/d\log\Mstel < 1$. {\it Why} the (high-mass) \MS\ slope is $<1$ is an interesting question, but it is unclear if the details of downsizing will be edifying in that context.

We discuss other such aspects related to the structure/``genetics'' of the G13 model in the next Section.

\subsection{Exploring G13's Structure}
\label{sec:genotype}

Section \ref{sec:phenotype} dealt with G13's appearance in frames defined by a range of cross-sectional data. This ``phenotype''---the outward manifestations of a suite of smooth, continuous, lognormal SFHs---compares favorably to most the observations studied, but because the model lacks explicit physical prescriptions, interpreting this fact is non-trivial.

To derive a physical interpretation, we now turn to G13's ``DNA''---the structural properties/axiomatic underpinnings/intrinsic features that lead to the above results. Our confidence in the model depends more on the accuracy of ideas in this Section than tests in the last.

\begin{figure}[t!]
	\centering
	\includegraphics[width = \linewidth, trim = 0cm 0cm 0cm 0cm]{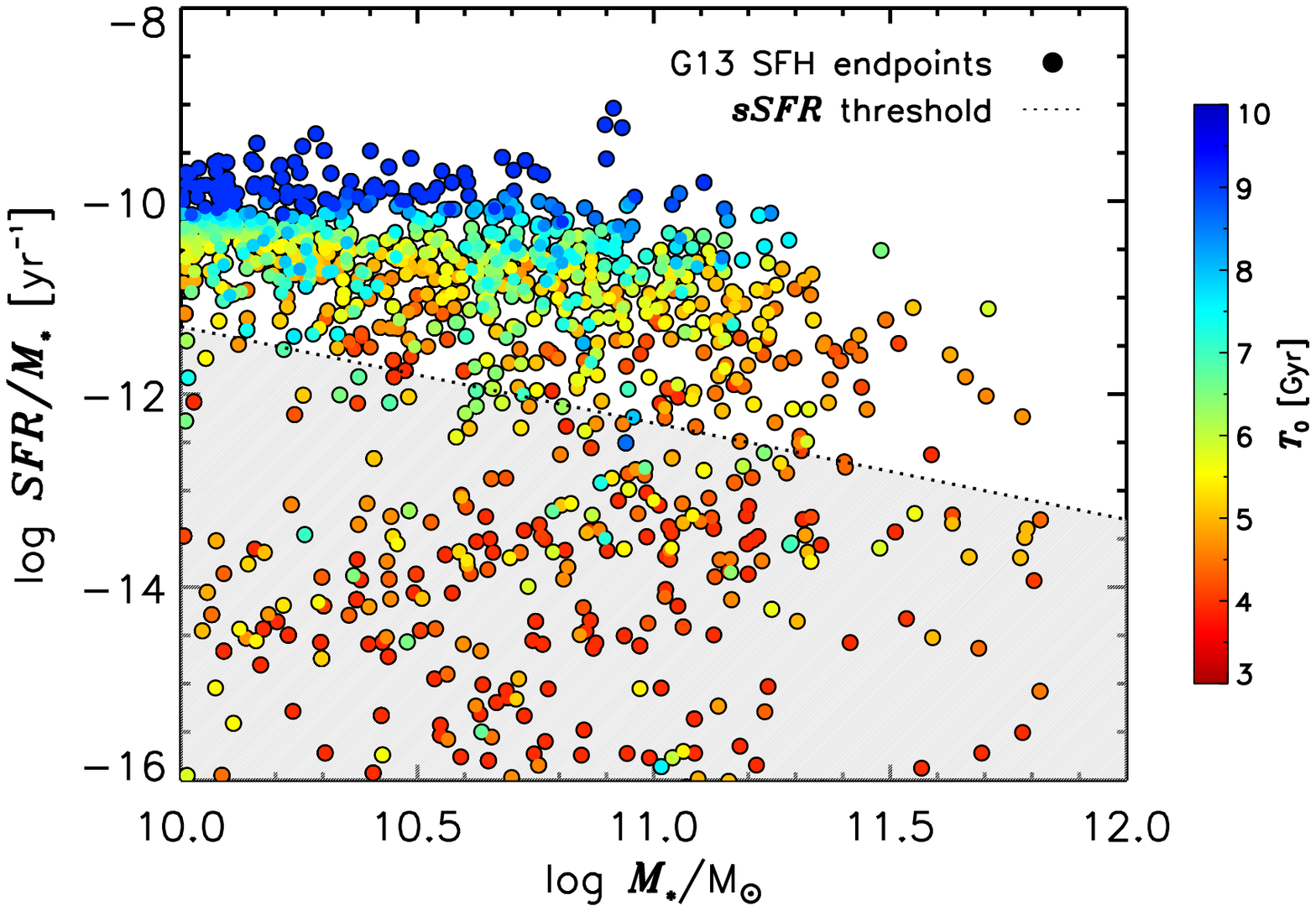}
	\caption{The G13 $z\approx0$ \MS\ colored by half-mass times, $T_{0}$. Gradients at fixed mass thus reflect {\it age} gradients. These persist at all $\Mstel$ over many epochs (Figure \ref{fig:SFMStrends}) and so are key to the model's interpretation of the \MS. They need not have emerged and {\it cannot} be imposed without violating other constraints (O13). The dotted line shows the SFR limit ($0.05\, \Msun\, {\rm yr^{-1}}$; see G13) above or below which the model reproduces a galaxy's measured $\Mstel$ {\it and} $\sfr$, or measured $\Mstel$ but arbitrarily low $\sfr$, \resp.}
\label{fig:todaySFMS}
\end{figure}

\begin{figure*}[t!]
	\centering
	\includegraphics[width = 0.75\linewidth, trim = 1cm 0cm 1cm 0cm]{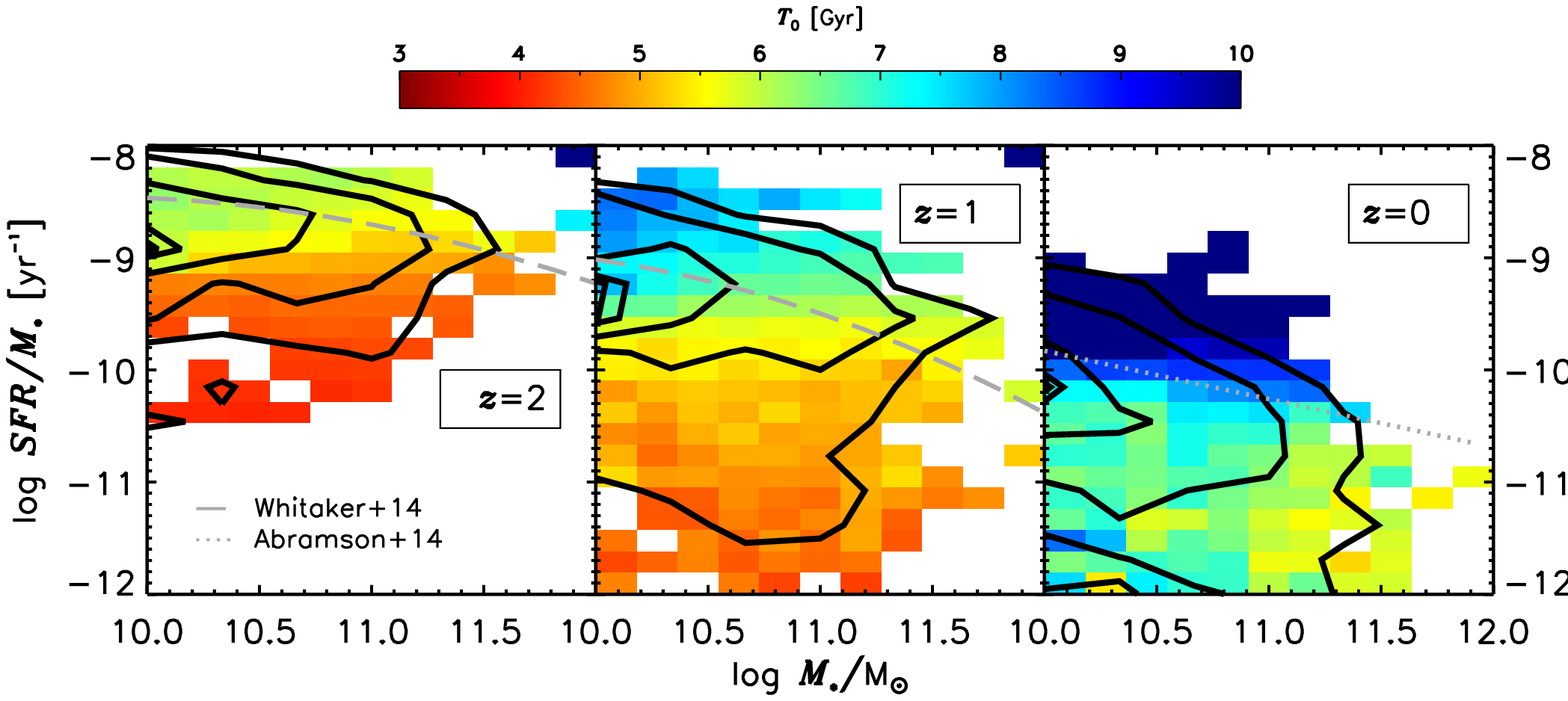}
	\includegraphics[width = 0.75\linewidth, trim = 1cm 0cm 1cm 0cm]{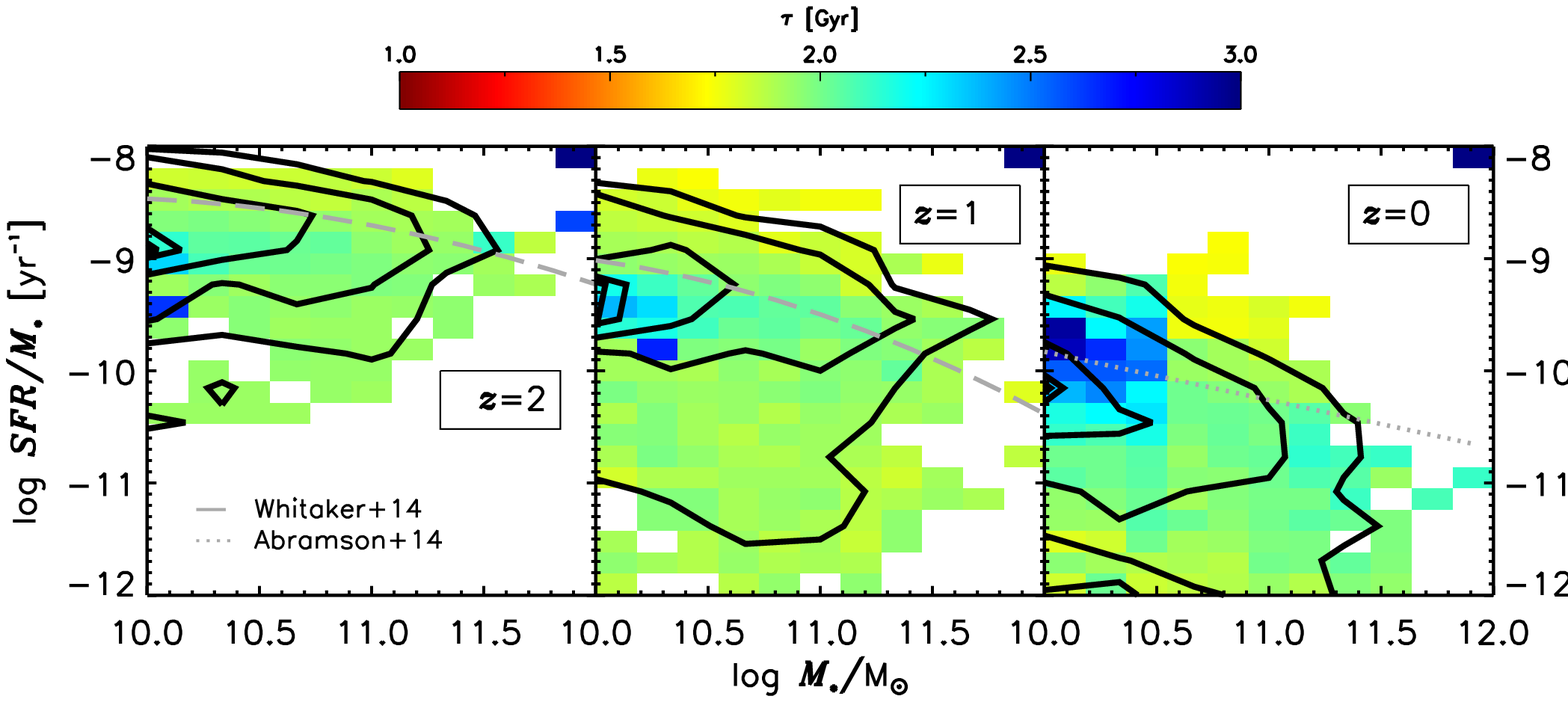}
	\caption{{\it From left}: G13 model projections onto the $\ssfr$--$\Mstel$ plane at $z \in\{2, 1, 0\}$. Density contours---5, 16, 50, 84, 95\% of max---roughly highlight the \MS. For comparison's sake, \MS\ fits from \citet[][dashed]{Whitaker14} and \citet[][dotted line]{Abramson14a} are overplotted in grey. Colors in the top and bottom panels reflect the mean $T_{0}$ and $\tau$ values of SFHs in 0.2$\times$0.2 dex boxes, \resp. Rainbows at fixed $\Mstel$ ({\it top}) reveal that the age gradients seen in Figure \ref{fig:todaySFMS} (top-right panel here) persist to $z=2$, implying that long-term SFH differentiation is {\it and has always been} a major source of the \MS\ scatter. Interestingly, $\tau$ shows no such cross-sectional stratification, suggesting that SF {\it timescale} is not a source of $\sigms$. However, this apparent decorrelation is due to the mixing on the \MS\ of SFHs terminating at different $\Mstel(z=0)$ at all $\Mstel(t)$. This washes out real longitudinal trends in $\tau$ that are critical to diversifying true evolutionary lineages (Section \ref{sec:newSpaces}; Figures \ref{fig:delSFMS}, \ref{fig:delTau}, \ref{fig:BTstuff}). An 0.3 dex cross-calibration offset has been applied to the G13 $\ssfr$s in all panels (see Figure \ref{fig:schema}). This Figure is analogous to Figure 2 of \citet{Tacchella16}, {\it left}.}
\label{fig:SFMStrends}
\end{figure*}

\subsubsection{The G13 Interpretation of the {\rm \MS} Scatter}
\label{sec:sigms}

The most salient structural aspect of G13 is its interpretation of the \MS. As illustrated by Figure \ref{fig:todaySFMS}, it reads $\sigms$---the scatter in (s)SFRs at fixed $\Mstel$---primarily as a gradient in the lognormal half-mass-time parameter, $T_{0}$. That is, {\it the scatter in the \MS\ reflects an age gradient in this model}. 

Indeed, the model suggests that even the ``small'' observed dispersion of $\sim$0.3 dex can accommodate a $\sim$7 Gyr spread in half-mass times. Intriguingly, using a totally different approach, \citet{Speagle14} derive a  similar interpretation for $\sigms$, and an almost identical $\sim$6 Gyr ($2$-$\sigma$) spread. As with K14, however (Section \ref{sec:meanSSFR}), the models differ meaningfully in how $\sigms$ is generated/maintained (see below).

We stress that no aspect of the modeling procedure explicitly induces such gradients. Instead this behavior is emergent; the addition of $\ssfr$ constraints at any number of epochs likely cannot alter it.

This fact is illustrated by Figure \ref{fig:SFMStrends} ({\it top}), which shows that the age gradients are maintained to $z=2$. Indeed, they persist to $z\gtrsim4$ (not shown), where horizontal stratification begins to dominate over the vertical structure illustrated here.

\begin{figure*}[t!]
	\centering
	\includegraphics[width = \linewidth, trim = 1cm 0.5cm 1cm 0cm]{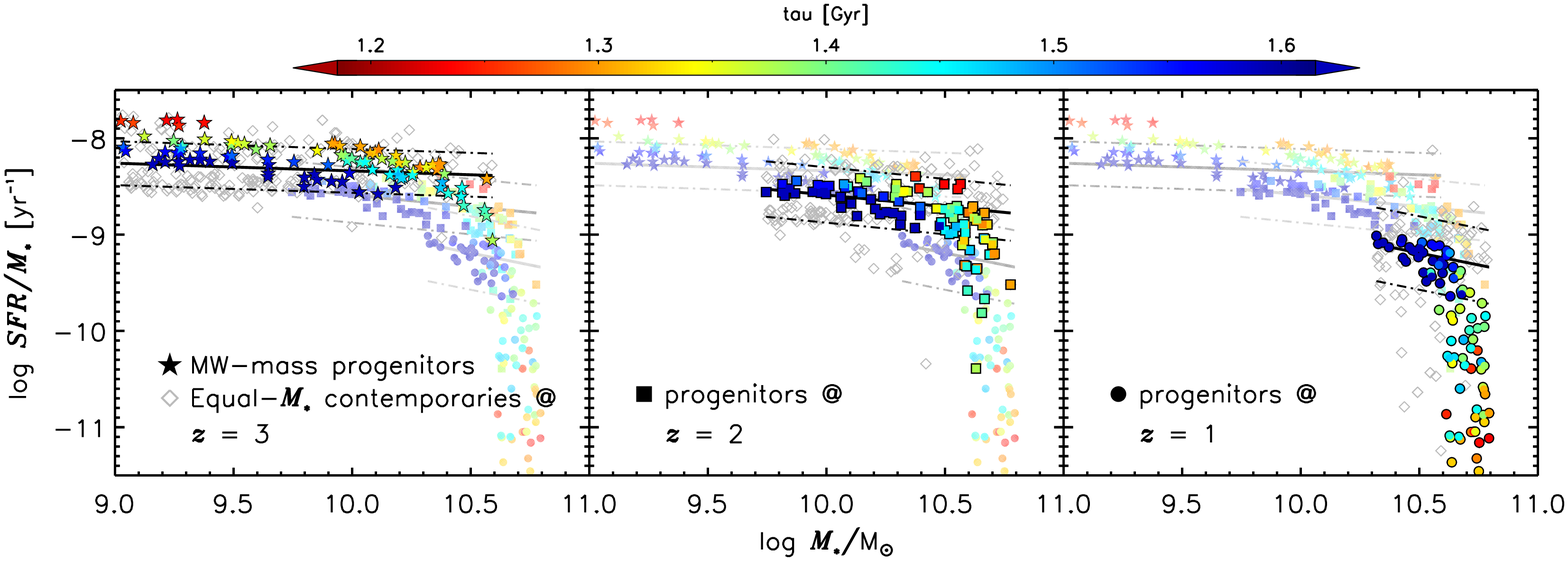}
	\caption{{\it Colored points}: G13 predictions for the locations of galaxies that actually evolve into today's MW-mass systems in $\ssfr$--$\Mstel$ space at $z=3$, 2, and 1 (from {\it left}; stars, squares, circles, \resp). Color coding reflects each SFH's $\tau$ value. In this {\it longitudinal} view, a clear relationship emerges between $\tau$ and $\ssfr$: galaxies with compressed histories (small-$\tau$) start high on the \MS\ and then fall below it, while galaxies with extended histories (large-$\tau$) remain at or slightly below its midline. This correlation---reminiscent of observed and simulated trends with gas and/or stellar surface density \citep{Ostriker11, Genzel15, Tacchella16, Zolotov15}---vanishes when the full \MS\ is plotted (Figure \ref{fig:SFMStrends}, {\it bottom}). This is because galaxies that {\it will not evolve into MW-mass systems} cohabit the locus with true progenitors at all $\Mstel$ (grey diamonds). Those systems display different $\tau$ trends depending on their destination, so mixing them with true MW-mass progenitors washes-out the signal shown by the colored points here. We replot all loci in all panels (faded) to show the full time evolution of the SFHs/\MS\ [cf.\ \citet{Tacchella16}, Figures 2 ({\it right}) and 7].}
\label{fig:delSFMS}
\end{figure*}

Two points are notable: 

First, the appearance of $T_{0}$ gradients means that G13 sees $\sigms$ as encoding {\it long-term differentiation} of galaxy SFHs. That is, $\sigms$ is not (or at least need not be) due to short-timescale, $t \ll t_{\rm Hubble}$, ``weather-like'' variations in SFRs. Instead, as in \citet{Speagle14}, even starforming galaxies at fixed $\Mstel(t)$ have---and have always had---a wide range of histories: Some reached half that mass many Gyr ago, others quite close to the epoch of observation. Thus, {\it equal-mass galaxies on the \MS\ need not form an evolutionary cohort.} Figure \ref{fig:delSFMS} (detailed below) reinforces this claim.

This finding accords qualitatively and quantitatively with new SFH measurements by \citet{Dressler16}. It is also similar to hydrodynamic numerical results from \citet{Tacchella16} showing the \MS\ scatter to reflect $\sfr$ variations of $\sim 0.5\, t_{\rm Hubble}$ (albeit at epochs when $t_{\rm Hubble}$ is only a few Gyr). We return to that work in Section \ref{sec:moreTau}.

Of course, in reality, starbursts and observational errors contribute to $\sigms$. Hence, G13 yields something of an upper-bound to the amount of differentiation that quantity encodes, and thus the timescale of SFH segregation. Regardless, G13-like models suggest that {\it a significant portion of the \MS\ scatter reflects galaxy diversification on Hubble-like timescales} (as seen by \citealt{Dressler16}), and is therefore a manifestation of a mechanism operating globally and probably at very early times (e.g., initial conditions; Section \ref{sec:ICs}).

The second point is that while G13 sets the \MS\ scatter mostly as an age/half-mass-time gradient, it does not do so exclusively: There are late-forming galaxies lying well below the \MS\ in Figure \ref{fig:todaySFMS}, and early-forming ones lying on it (though none at the very top). This subtle-but-real violation of the age-gradient interpretation---i.e., the breaking of age/mass rank-ordering---is critical, and where G13 departs from \citet{Speagle14}.

The SFHs that most dramatically drive this phenomenon are precisely those posited by O13 and subsequently spectrophotometrically identified by \citet{Dressler16}. They correspond to the $\sim25\%$ of galaxies with present-day $\Mstel\gtrsim4\times10^{10}\,\Msun$ that had late-rising {\it but also narrow} SFHs. That is, SFHs that rose and fell {\it faster} than the \MS. 

Single-parameter SFH models cannot describe this behavior: they cannot attain sufficiently high $\ssfr$s ($>1/t$) at $z\sim1$ and low ones today. To meet the first criterion, they must be constant or late{\it -and-continually} rising, automatically precluding them from meeting the second (O13 Figure 1). The same mathematics implies that $\sigms$ could {\it only} reflect an age gradient in a universe containing only those SFH forms \citep[or ones confining galaxies to the \MS;][]{Speagle14}.

O13 showed that we live largely---but not entirely---in that universe. To make up the difference, a two-parameter form with independent characteristic times and timescales was required. Lognormals fit this bill (through $T_{0}$ and $\tau$, \resp) with significant physical implications: The resultant ability for SFHs to fall faster than the \MS\ translates to allowing galaxies to exit the starforming population {\it without imposing explicit quenching}. Hence, this feature is one of the model's core axioms (Appendix \ref{sec:AA}), and the fact that it produces the general trend predicted by single-parameter SFHs (accurate for many galaxies) but also known violations \citep{Dressler16} is one of its key---and perhaps most informative---successes. 

The model cannot speak to the environments of the galaxies to which late-peaking, narrow (i.e., {\it young}) SFHs are ascribed, but O13 showed that they are probably isolated. \citet{Dressler16} later verified that that, while present, their abundances decrease in dense regions, complementing \citet{Poggianti13a}'s finding of higher fractions of older objects in clusters. These results suggest that SFH diversity is fundamentally related to the (initial) configuration of the matter density field. This seems compatible with \LCDM\ structure growth, where a halo's initial density/concentration maps closely to its mass accretion history \citep[e.g.,][]{Lacey93, Avila-Reese98, Lemson99, Wechsler02, Diemer15}. Section \ref{sec:discussion} revisits this point.

\subsubsection{What about $\tau$?}
\label{sec:tau}

As just mentioned, $T_{0}$---the lognormal half-mass-time---is important to generating $\sigms$ in the G13 scheme. Yet, as also just mentioned, a key feature of the model is that it treats $\tau$---the SFH width---independently. So, what does it do?

Interestingly, the bottom panel of Figure \ref{fig:SFMStrends} shows almost no correlation between $\tau$ and $\ssfr$ at fixed stellar mass. Thus, the G13 model suggests that formation {\it timescale}---unlike formation time---does not play a large role in generating $\sigms$, at least not when the locus is examined in the frame of the observer (Section \ref{sec:newSpaces}).

This behavior is interesting because correlations between, e.g., galaxy structural parameters and their location on the \MS\ at fixed $\Mstel$ appear weak in some works \citep[][but see Figure \ref{fig:BTstuff}, below]{Wuyts11,Bluck14,Tacchella16}.\footnote{This is not true of gas properties such as SFR surface density \citep[e.g.,][]{Wuyts11, Nelson15, Tacchella16}, or when considering galaxies of all $\ssfr$s \citep[e.g.,][]{Omand14}.} Hence, ironically, the lack of a strong gradient in $\tau$ at fixed $\Mstel$ across the \MS\ suggests that it, and not $T_{0}$, might best-link the G13 SFHs to observed galaxy structural properties. We explore this idea and the consequences of moving out of the observer's frame in the next Section.

\subsection{New Observational Spaces}
\label{sec:newSpaces}

So far, we have studied how the G13 model expresses itself in the context of canonical observations in galaxy evolution. Even when discussing its structure, we have shown projections into observed frames.

This approach is necessary to validate the model. Yet, because it is longitudinal---containing information not present in the data on the (theoretical) time-evolution of individual galaxies (Section \ref{sec:glossary})---we need not limit ourselves to cross-sectional snapshots---to ``connecting the dots'' between different systems at different epochs---to learn from G13. Indeed, we need not limit ourselves to data at all.

In this Section, we first show how moving out of the frame of the data reveals new physical insights. We then apply those insights to a new space---galaxy bulge mass fractions---and obtain results remarkably in-line with observations. If these exercises do not illustrate successes of the G13 model, they provide good avenues by which to assess its ultimate validity.

\subsubsection{A Longitudinal View of $\tau$}
\label{sec:moreTau}

Figure \ref{fig:SFMStrends} ({\it bottom}) shows no $\tau$ trends at fixed mass along the \MS. Yet, this presentation has limited virtue in elucidating how $\tau$---and therefore what physics---differentiates galaxies because not all equal-$\Mstel$ galaxies on the \MS\ are evolving towards the same destination (Section \ref{sec:sigms}). When galaxies that {\it will} ultimately evolve to a similar final mass are selected {\it a priori}, the G13 $\tau$ trends change dramatically.

Figure \ref{fig:delSFMS} shows the evolution of all SFHs terminating near the Milky Way's (MW's) stellar mass ($10.6\leq\log\Mstel\leq10.8$) that had $\log\Mstel(z=3)\geq9$ (to simplify the illustration). At early times, there is a clear preference for systems with low $\tau$---i.e., narrow SFHs---to lie high on the \MS, while those with high $\tau$---elongated SFHs---lie low. At late times, however, this trend reverses, such that low-$\tau$ galaxies lie well below their high-$\tau$ counterparts. So, if such true progenitors could be isolated, what appeared as a decorrelation in cross-sectional data would transform into a clear trend.

On reflection, this trend {\it must} be present (Section \ref{sec:GV}, Appendix \ref{sec:AC}): If a galaxy is to reach a given final mass before others do, it must grow rapidly early on---surging ahead of its eventual peers (middle panel)---and then stop, moving from a high to low position on (and then off) the \MS\ (consistent with the findings of \citealt{Marchesini14}). However, this trend is not apparent {\it in the data} because galaxies that will ultimately evolve into systems of different final masses commingle with the true evolutionary cohort and obliterate this signal (grey diamonds in Figure \ref{fig:delSFMS}).\footnote{The \MS\ is a train: people in different cars get off at the same station, people in the same car get off at different stations. Figure \ref{fig:SFMStrends} ({\it bottom}) tracks the train and reveals a $\tau$ jumble. Figure \ref{fig:delSFMS} tracks only passengers going to the same stop and this reveals a meaningful trend. If a third parameter were found that mapped better to ``time on train'' than $\Mstel(t)$, $\tau$ should correlate with it in a 3D version of Figure \ref{fig:SFMStrends}.}

Note, however, how much real estate is covered by what will ultimately be galaxies of comparable stellar mass: At $z=3$, such systems span {\it more than a factor of 30} in that quantity. This number is not inconsistent with theoretical results based on abundance-matching. Using the tool supplied by \citet{Behroozi13a}, we find the 2-$\sigma$ spread of {\it halo} masses to be a factor of $\sim 16$ for the $z=3$ progenitors of today's MW-mass halos [$\Mhalo(z=0)\simeq10^{12}\,\Msun$; e.g., \citealt{Watkins10}]. This grows to a factor of $\sim18$ in {\it stellar} mass assuming a $0.2$ dex scatter in $\Mstel(\Mhalo)$ \citep{Behroozi13}. As such, the factor of $\sim30$ suggested by the model seems not so large as to to rule it out on number-density conservation grounds (i.e., by requiring a wildly high merger rate).

Regardless, Figure \ref{fig:delSFMS} depicts a simple encapsulation of G13's core astrophysical proposition: {\it the interesting physics in galaxy evolution is revealed by contractions of factors of 30 in $\Mstel$ to eventual factors of $<2$}. Because mean-based techniques such as \MS\ integration \citep{PengLilly10, Leitner12} average over {\it that} process, these approaches may neglect critical parts of the story, blinding themselves to sources of galaxy diversity beyond quenching. In the extreme, the G13 model suggests that there is little-to-no astrophysical utility in describing the average SFH of galaxies at fixed $\Mstel(t)$.

A final note in this vein: Examining a plot similar to Figure \ref{fig:delSFMS} but color-coded by $T_{0}$ shows less (or no) vertical stratification, but instead greater differentiation along the mass axis. This trend is also expected---more-massive galaxies at any epoch should have earlier half-mass times almost by definition---but it is {\it orthogonal} to that seen in the full G13 \MS\ (Figures \ref{fig:todaySFMS}, \ref{fig:SFMStrends} {\it top}), again emphasizing the limitations of that locus as a window on evolutionary lineages.

\subsubsection{Towards Physics}
\label{sec:toPhysics}

\begin{figure*}[t!]
	\centering
	\includegraphics[width = \linewidth, trim = 1cm 0.5cm 1cm 0cm]{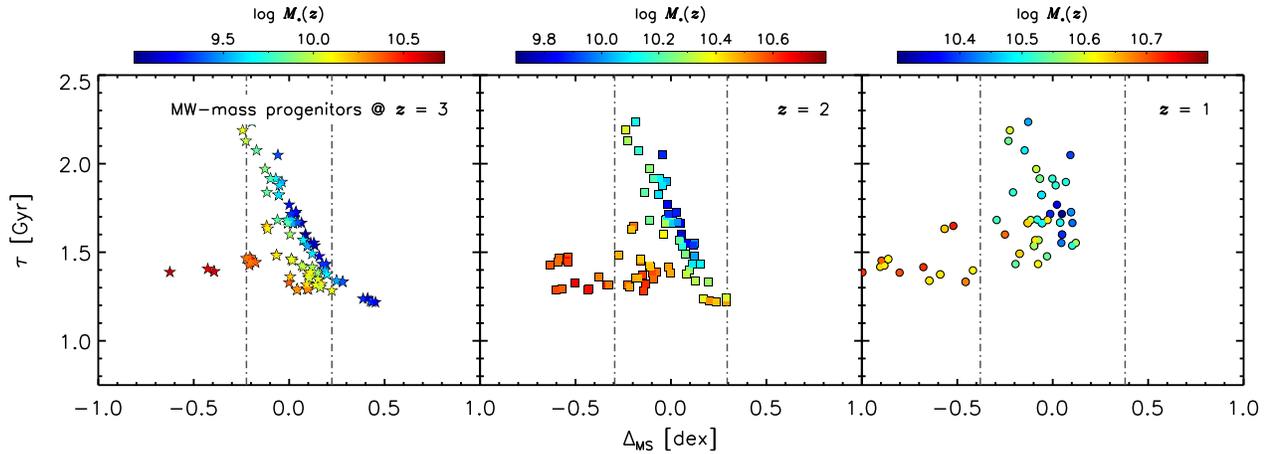}
	\caption{The relationship between $\tau$ and distance from the \MS\ midline ($\Delta_{\rm MS}$) at the epochs in Figure \ref{fig:delSFMS}. Points correspond to that Figure but color-coding reflects $\Mstel$ here, with redder points being closer to their $z=0$ mass. This is the mathematical definition of quenching, but such systems also lie towards the left (underproductive) region of the diagram---the observational definition of quenching. Conversely, bluer points, which will grow substantially in the interim, span all positions on the \MS. Their $\tau$ values tend to an anti-correlate with $\Delta_{\rm MS}$ such that more rapidly growing (low-$\tau$) starforming systems lie high while more slowly growing (high-$\tau$) systems lie low. This trend is similar to those seen when $\tau$ is replaced by (inverse) \Sersic\ index or SF/stellar mass surface density [e.g., \citealt{Schiminovich07} (cf.\ their Figure 17), \citealt{Wuyts11} (cf.\ their Figure 4)], or gas depletion timescale \citep[][cf.~their Figure 7]{Tacchella16}. Combined with results shown below (Figure \ref{fig:BTstuff}), this suggests $\tau$ encodes a density-like property. Since systems do not evolve in $\tau$ (i.e., along the $y$-axis) in G13, a prime candidate is the {\it initial} (baryonic) density in some physical aperture defined at a galaxy's birth.}
\label{fig:delTau}
\end{figure*}

When examined from the perspective of galaxies that will evolve into similar-mass systems, Figure \ref{fig:delSFMS} shows $\tau$ to be a potentially important hook for physics. Figure \ref{fig:delTau} attempts to clarify this by moving into the frame of the \MS. Here, we show the offset of each SFH from the \MS\ midline as a function of $\tau$ at $z\in\{3,2,1\}$. The points are coded now by $\Mstel(z)$, revealing how close they are to ``quenched'' (more-properly ``finished,'' since the mathematical condition is $\Mstel=\Mstel^{\rm final}$) at any epoch.

The trend described in Section \ref{sec:moreTau} is more obvious here: galaxies with extended SFHs---large $\tau$---sit near the mean \MS\ ($\Delta_{\rm MS} = 0$), while those with narrow SFHs---small $\tau$---occupy all positions across it with $\Mstel$ being the determining factor. The tail of low-$\tau$ high-$\Mstel$ galaxies drifting below the \MS\ (to the left in Figure \ref{fig:delTau}) at any epoch might be seen as quenching, but G13 suggests it is better seen as a direct consequence of the fact that, if a galaxy is to reach its final mass early, it must have a compressed SFH \citep[][]{Wellons15}.

The most interesting objects are those with low-$\tau$ {\it and} low-$\Mstel$ (bottommost blue points in Figure \ref{fig:delTau}). These objects are rapidly shooting across the top of the \MS\ in Figure \ref{fig:delSFMS}. This behavior is potentially instructive.

If we make the reasonable assumption that $\tau$---the ``span'' of the SFH---to some extent reflects gas consumption timescales, which are perhaps {\it set} by gas surface densities \citep[or, in a globally averaged sense, filling factors][]{Kennicutt98, Bigiel08, Genzel10, Ostriker11}, this Figure is evocative of the findings of \citet{Genzel15}, \citet{Tacchella16}, and compatible with those of \citet[][see their Figures 12, 13]{Zolotov15}. That is, galaxies with compressed SFHs are those that consume gas rapidly perhaps {\it because} it is in an extremely dense configuration. 

This scenario, which echoes \citet[][see his Figure 6]{Holmberg64}, naturally explains trends from, e.g., \citet{Franx08}, \citet{Williams10}, \citet{Valentinuzzi10a}, \citet{Fang13}, \citet{Omand14}, \citet{Barro15}, and \citet{Morishita16b}: Dense {\it starforming} galaxies evolve rapidly, so, at any epoch, galaxies that appear to be quenching will be dense, leaving dense stellar configurations behind to be preferentially identified as passive galaxies in a future epoch. However, because G13 contains no such physics, {\it galaxies do not evolve towards denser states in this model}. As such, the above picture need not entail a ``compaction'' event (though it could; \citealt{Zolotov15, Barro15}). Instead, Figures \ref{fig:delSFMS} and \ref{fig:delTau} suggest quenching is the ``peeling-off'' of the densest tail of the starforming population at any epoch, explaining why the ($\Mstel$ and) surface density threshold of \citet[][see their Figures 1, 7]{Barro15} falls with time.

Ultimately, both scenarios---compaction, in which SF galaxies evolve towards a denser state before quenching, and ``peeling,'' in which the densest SF galaxies simply ``finish first''---may be correct. Our fundamental argument is that, because the model has no density parameter, yet captures something like the density-related trends seen in data through $\tau$, {\it the fundamental physics of interest is whatever sets the $\tau$ spectrum}: it is {\it that} which determines if and when a galaxy will undergo either of the above processes.

An attempt below to tie G13 to data similarly outside its formal scope supports this interpretation.

\subsubsection{Bulges and Disks}
\label{sec:BT}

G13's stance on $\tau$ is that it should appear unrelated to a galaxy's position on the \MS\ in cross-sectional data---i.e., observations---but should still be important to diversifying {\it actual members} of an evolutionary lineage. If we could identify an observable that maps to $\tau$ we could define cuts in, e.g., the $\ssfr$--$\Mstel$ plane that isolate true progenitors but are otherwise invisible in the data.  

\begin{figure*}[b!]
	\centering
	\includegraphics[width = 0.8\linewidth, trim = 1cm 1.5cm 1cm 0cm]{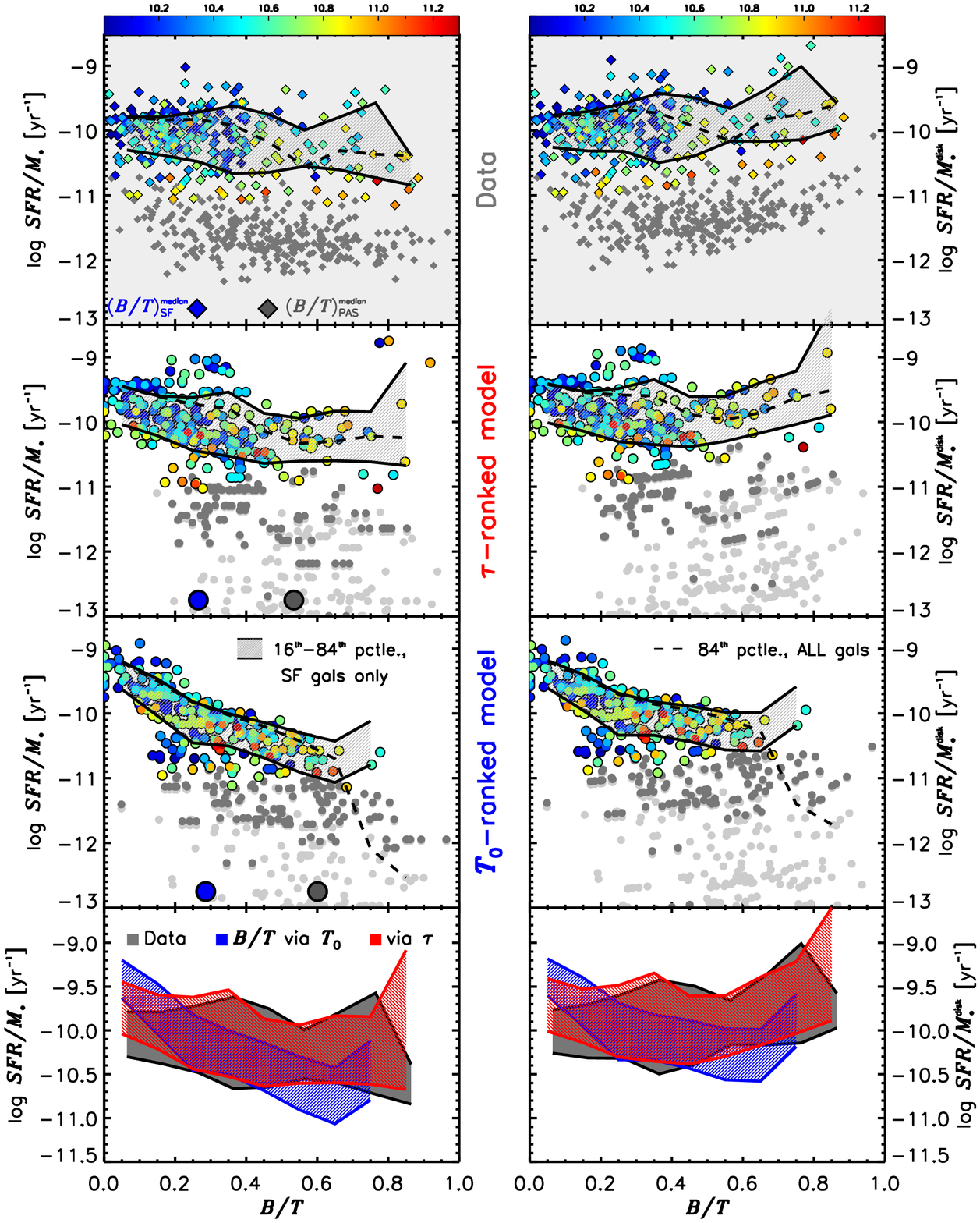}
	\caption{Projections of the $z=0$ $\sfr/\Mstel$-- and $\sfr/\Mstel^{\rm disk}$--$B/T$--$\Mstel$ planes (left/right columns, \resp). Data are shown at {\it top} (grey panels; \citealt{Brinchmann04,Gadotti09}). Recreations using $\tau$ or $T_{0}$ to assign $B/T$s from the top panel to the G13 SFHs are shown in the second and third rows, \resp. Colored and grey points show ``starforming'' and ``non-starforming'' galaxies/SFHs, \resp\ ($\sfr<$ SFMS $-$ 2-$\sigms$; light grey points show G13 SFHs with $\sfr(z_{\rm obs})<0.05\,\Msun\,\yr^{-1}$; Equation \ref{eqn:sfrCons}). Shaded regions highlight the 1-$\sigma$ spread of the starforming systems only. Dashed lines in the top six panels show the $84^{\rm th}$ percentile of all galaxies (including $\ssfr$ upper-limits; attempts to capture the ``shower rod'' from which the ``curtain'' of galaxies descends). Point colors reflect present-day $\Mstel$. As summarized in the bottom panels, $\tau$-ranking---assigning the narrowest G13 SFH per mass-bin to the highest $B/T$, and vice-versa---reproduces the data trends far better than $T_{0}$-ranking---where the earliest-forming G13 SFH gets the highest $B/T$. This is true in terms of (1) the slope and width of the $\sfr/\Mstel^{\rm (disk)}$--$B/T$ relation for SF galaxies; (2) the maximal $B/T$ of SF galaxies; (3) the location of the upper-envelope of {\it all} galaxies; and (4) the median $B/T$ for the SF and non-SF populations (blue/grey symbols, \resp, on the top three $x$-axes at {\it left}).}
\label{fig:BTstuff}
\end{figure*}

Oddly, the lack of an apparent correlation of star formation activity with $\tau$ in cross-sectional samples may hint at which observable(s) it is linked to. For example, at similar masses, \citet{Wuyts11} showed that trends in galaxy \Sersic\ indices are also weak at fixed $\Mstel$ across most of the \MS. \citet{Fang13} and \citet{Barro15} (mentioned above) observed something similar for stellar mass surface density, which \citet{Tacchella16} also see in simulations. These weak correlations are well-reflected by $\tau$'s behavior across the \MS\ (Figure \ref{fig:SFMStrends}, {\it bottom}), but not $T_{0}$'s, which shows strong gradients at fixed mass (Figure \ref{fig:SFMStrends}, {\it top}).

Going one step further, if we associate \Sersic\ index or stellar mass surface density with bulge mass fractions, $B/T$, then the connection posited above between $\tau$ and a gas consumption or compaction timescale would also encourage us to link $\tau$ to $B/T$, since compact galaxies with short formation timescales likely evolve into bulge-dominated systems \citep{Schiminovich07, Barro13, Barro15, Steinhardt14, Zolotov15, Tacchella16}.

We test this association in Figure \ref{fig:BTstuff}. This Figure is complex, so we review it in some detail.\footnote{\citet[][]{Lilly16} provide an alternative test to what follows. While they work from an independent perspective, there is nothing obviously incompatible between our approaches or conclusions.}

The first step in its construction is to assign each G13 SFH a $B/T$ ratio. We do this in a manner similar to the ``age matching'' prescription of \citet{Hearin13}. 

Using a sample of high-quality $z\approx0$ $B/T$ measurements \citep{Gadotti09} and moving in bins of $\Mstel$, we rank-order the data by $B/T$ and the SFHs by either $\tau$ (second row) or $T_{0}$ (third row). We then simply map the lowest-$\tau$ or lowest-$T_{0}$ SFH to the {\it highest} $B/T$ and continue down both lists. Physically, this corresponds to assuming either rapidly (low-$\tau$)  or early forming galaxies (low-$T_{0}$) become bulge-dominated, while slowly/late forming ones become disk-dominated.

We do this without respect to SFR. As such, the known $z=0$ $B/T$--$\Mstel$ relation---slope and scatter---is reproduced by construction, but the SF properties of high- and low-$B/T$ systems is not. Examining either the trend of $\ssfr$ ($\sfr/\Mstel$) or disk-mass normalized $\sfr$ ($\sfr/\Mdisk$) with $B/T$ (left and right columns, \resp) may therefore yield hints as to which G13 model parameter is most closely associated with bulge growth---or stellar surface density, or \Sersic\ index, or ``compaction''---and enhance our physical interpretation.

The top row in Figure \ref{fig:BTstuff} (grey panels) shows the data using aperture-corrected $\sfr$s from \citet{Brinchmann04} (DR7 release). Colored points (coded by $\Mstel$) are ``starforming'' galaxies defined by a 2-$\sigma$ cut below the \MS. As expected, high-mass systems have reduced $\sfr/\Mstel$ compared to low-mass systems, and high-$B/T$ systems have reduced $\ssfr$ compared to low-$B/T$ ones (highlighted by the grey shaded bands, showing the 16th--84th percentile spread of the starforming population). As also anticipated \citep{Abramson14a}, high-mass, high-$B/T$ systems have $\sfr/\Mdisk$ more comparable to lower-mass or diskier systems. Note that the distribution of either quantity at fixed $B/T$ is broad.

Comparing the data to the model shows that $T_{0}$-ranking (third row) poorly reproduces these trends; both $\sfr/\Mstel$ and $\sfr/\Mdisk$ fall too steeply with $B/T$, and loci at fixed $B/T$ are too narrow. The former is true considering only starforming galaxies, or the upper-limb of all galaxies (dashed black lines in all panels). Further, the median $B/T$ for {\it non}-starforming galaxies is $\sim15\%$ higher than that in the data (large symbols near abscissas). 

In all respects, $\tau$-ranking (second row) does a far better job. As summarized in the bottom row of Figure \ref{fig:BTstuff}, the slopes and widths of the observed trends for starforming (or all) galaxies are much more faithfully reproduced using this method, as is the median non-starforming $B/T$ ratio.

This outcome is precisely what is expected if $\tau$ is indeed linked to gas consumption timescales, or gas/stellar mass surface densities as Figures \ref{fig:delSFMS} and \ref{fig:delTau} suggest. As such, it deepens our sense that $\tau$ encodes the action of a key organizational principle in galaxy evolution.

Note: Figure \ref{fig:BTstuff} and the above discussion are not intended to convince the reader that there is, in reality, a 1:1 mapping between $\tau$ and a galaxy's present-day $B/T$. Neither are we suggesting that we have found---or even attempted to find---an accurate description of the data using the G13 model. We {\it are} suggesting, however, that a model with no physics combined with very simple prescriptions to extend it to domains quite far from anything it was intended to describe can yield highly suggestive and potentially informative results. 

We believe this fact alone is weighty. However, the results in Figure \ref{fig:BTstuff} combined with the phenomenological resemblance of $\tau$'s relationship to the \MS\ and that of the above phenomena embolden us to posit that we are not looking at a mere description of the data in these cases, but in fact an explanation. That is, $\tau$---or some optimal combination of $T_{0}$ and $\tau$ we have yet to discover---is not only correlated with, but in fact {\it controls} the eventual gas/stellar surface density, bulge fraction, or size of a galaxy, and therefore its probability of being ``quenched'' at any epoch/as a function of time. As such, we will use what we have learned to make a prediction.

\begin{figure}[t!]
	\centering
	\includegraphics[width = \linewidth, trim = 1.2cm 0.5cm 1cm 0cm]{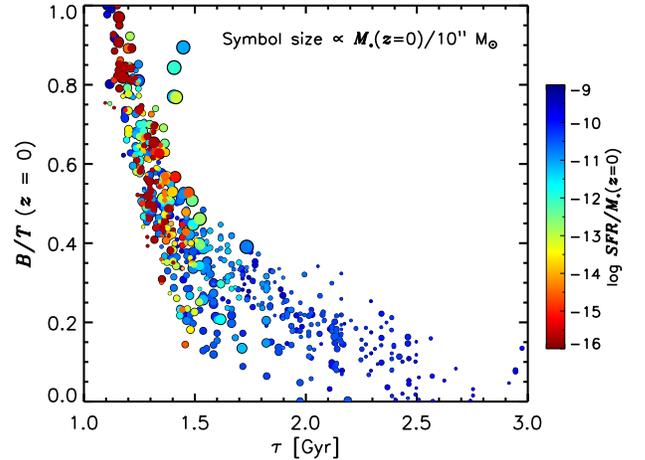}
	\caption{The present-day $\tau\mapsto B/T$ mapping that produces the second row of Figure \ref{fig:BTstuff}. Symbol colors and sizes reflect each SFH's $z=0$ $\ssfr$ and $\Mstel$, \resp. The $B/T$ assignment guarantees the correct symbol size--$y$-axis location relationship, but color-highlighted trends are emergent. These reveal G13 to reproduce known tendencies for starforming {\it and} passive galaxies to span a range of $B/T$, but for the former be disky and the latter almost free of pure disks (\citealt{Schiminovich07}, Oemler et al., in preparation), implying that short formation timescales are key to bulge-building.}
\label{fig:prediction}
\end{figure}

Figure \ref{fig:prediction} shows the G13 $\tau \mapsto B/T$ mapping as functions of a galaxy's present-day stellar mass (symbol size) and $\ssfr$ (symbol color). As demanded, low-$\tau$ objects have high-$B/T$ and vice versa, but the shape of this relationship and its coloring are telling. The steepness at the low-$\tau$ end---dominated by more-massive, more-passive galaxies---is reminiscent of the spread in $B/T$s exhibited by early Hubble type objects \citep{Dressler80,Kelson98,Fabricant00}. Meanwhile, the shallower trend at higher-$\tau$ shows the general diskiness of all late-type starforming galaxies today (though they span many morphological subclasses, perhaps reflecting a range of formation timescales). Furthermore, this mapping leads to {\it some} high-$B/T$ starforming galaxies, {\it but no} (very) low-$B/T$ passive ones (i.e., red points). This is true even though there are plenty of passive objects at masses dominated by pure-disks. These last subtleties are observed {\it facts} (e.g., \citealt{Schiminovich07}; Oemler et al., in preparation); that the model produces them along with the grosser trends mentioned above further supports its validity.

Since $\tau$ describes a formation timescale, it should also correlate with $\alpha$-element enhancement, for example---or, again, perhaps Hubble type. As such, we encourage other investigators to assemble such data and compare their observations to the predicted trend in Figure \ref{fig:prediction}. We also suggest that, now or in the (not-too-distant) future, advanced SPS modeling may allow reliable determinations of $\tau$. Hence, others might also wish to run such spectral/SED fitting using lognormal SFHs and compare their results to this diagram (or to assemble the relevant data when such technology is more robust). Finally, we would encourage numerical simulators to produce a similar diagram to provide an independent prediction from which deeper physical causes can perhaps be identified.

We finalize our physical interpretation of G13 below.


\section{Discussion}
\label{sec:discussion}

The phenomenological slog above was intended simply to persuade the reader that the G13 model---a collection of just 2094 lognormal SFHs---is sufficiently good at reproducing sufficiently many observations across sufficiently diverse domains and sufficiently long stretches of time that its deeper physical and methodological consequences merit serious consideration. There are three possible consequences:
\bitem
	\item[]{\bf Astrophysical} -- Galaxy growth trajectories can diverge radically from inferences based on the evolution of scaling laws (e.g.,  the \MS). They are not characterized by large discontinuities, more likely are  (largely) defined by initial conditions, and may (largely) share a common terminating force. {\bf Or...}
	\item[]{\bf Metaphysical} -- Current data can be described from orthogonal---if not opposite---astrophysical perspectives. Such disagreements must be settled or understood before tests of more explanatory models can be fully contextualized.  {\bf Or...}
	\item[]{\bf Epistemological} -- The information content of the observations examined above is (very) low.
\eitem
We now explore each of these statements in turn.

\subsection{Astrophysical Implications}
\label{sec:ICs}

\subsubsection{Quenching Is Uninformative}
\label{sec:noQuench}

For those who view the G13 model as a (sufficiently) good description of the data, our immediate conclusion is that quenching---defined any way that suggests the removal from the starforming population of galaxies with arbitrarily high $\ssfr$s---is a needlessly limiting physical framework, if not an overly mechanistic interpretation of the facts. We say this because G13 entails no quenching physics, and indeed no starforming ``states'' for galaxies to exist in at all---only a diversity of smooth SFHs evolving along independent trajectories that naturally rise and fall. 

This statement is broader than it might first appear. For example, it can be rephrased as ``Galaxies do not evolve {\it along} the \MS.'' As shown in Figure \ref{fig:delSFMS}, there is no need for starforming galaxies observed at fixed $\Mstel$ at one epoch to appear as a similarly confined cohort at any other epoch. The \MS\ therefore {\it shows you} galaxies evolving, but it does not {\it tell you} (much) about why they are doing it. In this sense, it and similar trends are like swarms of starlings: they emerge from flight paths of individual birds, but reveal few details about those trajectories (Section \ref{sec:descriptions}), and therefore little about how/why birds fly (or stay grounded). The histories tell the story; the loci do not. As such, the utility of inferring physics from the (evolution of) scaling relations is dubious. If galaxies are seen as evolving {\it through} the \MS, the need for an {\it ad hoc} mechanism to derail them from that trend is removed.

Our argument is slightly deeper than this. The invocation of grow-and-quench scenarios is natural given cross-sectional data: when galaxy properties at a single epoch are collapsed onto 2D planes, the obvious correlations that result encourage causal interpretations---or at least encourage us to think we are seeing evolutionary trends. Yet this cannot be true: starforming galaxies at a given epoch are the progenitors of precisely {\it none} of the passive galaxies at that epoch. Hence, such conclusions are valid if and only if quenching must be rapid, a scenario we and others have here and elsewhere shown need not be the case \citep[e.g.,][]{Dressler13,Schawinski14,Peng15}. If a distinction between longitudinal and (series of) cross-sectional samples is maintained, the need for/idea of quenching changes, or disappears.

One could argue at this point that the falling part of our histories {\it is} quenching. We would agree, so long as ``quenching'' means ``the smooth crossing of an arbitrarily low $\ssfr$ threshold''; i.e., {\it normal galaxy evolution}.\footnote{In some sense, we are suggesting that a universal null quenching scenario be adopted entailing Hubble-timescale processes---which are rapid at high-$z$---that should be rejected before other explanations are sought.} This might entail a combination of gas exhaustion and perhaps a mechanism to prevent some gas from condensing to fuel star formation \citep{Conroy15, Voit15}. The simplest interpretation of our model would be that the SFH declines---the phenomena requiring explanation---are due purely to the former: the gradual reduction in cold gas fractions over time \citep[][]{Geach11, Popping15}. However, as populations of non-starforming galaxies exist with non-negligible amounts of neutral gas \citep{Chen01, Thom12, Young14, Johnson15}, something akin to the latter is probably also at work. We emphasize that neither option is terribly exotic: although, e.g., supernovae and rapid AGN feedback must be important in some contexts, our findings suggest that they are generally sub-dominant to the above \citep[concurring with][]{Feldmann15}.

Again, one could argue that galaxies are known to suffer large discontinuities in star formation induced by, e.g., environmental gas stripping or starbursts \citep[e.g.,][]{GG72,DresslerGunn83,Poggianti16}, so ``fast-track'' quenching clearly exists. We admit that the current G13 model can neither capture nor constrain the prevalence of such events. However, we would argue that the results above---and the data \citep{Zabludoff96, Quintero04, Rodighiero11, Peng15, Dressler13, Abramson13, Dressler15}---suggest that these are perturbations to an underlying narrative based on smooth/long-timescale $\sfr(t)$ correlations, and are therefore corrections to a G13-like scheme.\footnote{We acknowledge that sufficiently narrow, smooth SFHs leading to red galaxies at $z\gtrsim2$ may be difficult to distinguish observationally from abrupt truncations of \MS-driven growth or powerful starbursts. Comparisons of the spatial distributions of galaxies binned by $\Mstel$ and $\sfr$ might distinguish these scenarios in the era of {\it WFIRST}, however. (see H.~Ferguson's talk at \url{https://conference.ipac.caltech.edu/wfirst2016/system/media_files/binaries/19/original/ferguson_wfirst2016.pdf?1457376009}).}

\subsubsection{Galaxy Evolution Is Driven by Initial Conditions}
\label{sec:natureIsNurture}

``Quenching is uninformative'' can be rephrased in yet another way: ``Environmental and internal effects may be inseparable'' \citep[see also][]{DeLucia12}. This is because our model makes no such distinction; it simply assigns each galaxy a $(T_{0},\tau)$ pair at birth. If these quantities---``clocks'' that set the speed of each galaxy's evolution---were somehow correlated on supergalactic scales early on, the causal effects of internal and external influences on SFHs would be linked. 

The immediate interpretation of the above \citep[shared by ][]{Speagle14} is that {\it initial conditions must substantially determine the course of a galaxy's evolution}. We believe that this is the fundamental implication of the G13 model, and that our conclusion regarding quenching is a consequence of it. If we are correct, from a paradigmatic standpoint, the most critical aspect of galaxy evolution is {\it which initial conditions} set (something like) {\it $T_{0}$ and $\tau$}.

An obvious candidate is the configuration of the matter density field, specifically the initial baryonic overdensity in some optimal aperture. This choice is motivated by Figures \ref{fig:todaySFMS}, \ref{fig:SFMStrends}, \ref{fig:delSFMS}, \ref{fig:delTau}, and \ref{fig:BTstuff}, in which $T_{0}$ and $\tau$ (mostly $\tau$) are seen to play roles similar to gas/SFR surface densities or gas consumption timescales as assessed by, e.g., \citet{Wuyts11}, \citet{Zolotov15}, and \citet{Tacchella16}. We believe that our model is also consistent with the work of \citet{Hearin13}, which suggests that a halo's age---the time at which it passes a mass, mass-accretion, or halo-membership threshold (see their Sections 4.1.2, 6.3)---is a critical parameter, in addition to, and separate from, its total mass. As it seems reasonable for halo baryon overdensity to be well-correlated with such a threshold, these may ultimately be two ways to describe the same phenomenon.

Regardless, the action of both theoretical mechanisms is substantiated by data, either in the densities of many starforming galaxies at high redshift \citep{Barro13,Barro15}, or the older stellar ages of passive galaxies at fixed $\Mstel$ in clusters compared to those in the field \citep{Thomas05,Poggianti13a,Mcdermid15,Morishita16b}, or the accelerated decline in the SFRD in overdense environments \citep{Guglielmo15}. 

To summarize, it is not that passive galaxies are ``quenched,'' but that they are ``finished,'' having had an accelerated assembly/formation history compared to starforming counterparts because they grew up in denser regions of the Universe. This scenario---which is not new, but a phenomenological rediscovery/synthesis of ideas from \citet{Holmberg64}, \citet{Tinsley68}, and \citet[][]{Dressler80}, and seems deeply related to the concept of assembly bias (e.g., \citealt{Wechsler06,Wetzel07,Dalal08,Feldmann16,Saito16}; and relatedly \citealt{AragonCalvo16})---naturally links environmental density, internal galactic density, and the characteristics of galaxies in overdensities as observed at effectively all epochs. {\it From that starting point, everything else emerges}. 

So, while it would be helpful to know if AGN-related activity ultimately produces the (spread of) SFH declines we have parametrized in a physics-agnostic manner, we would still ask if the action of that mechanism is a symptom of an underlying organizational principle such as the above, and suggest that confirming {\it that link} is important. Indeed, simulators may already have the information to do so: We would encourage them to see which aspects of their models set (parameters analogous to) $T_{0}$ and $\tau$, and whether identifying trends at that level might enhance their understanding of deeper physics.

\subsection{Metaphysical Implications: \\
Too Many Descriptions to Permit Explanations}
\label{sec:descriptions}

Issues will remain, however, independent of the outcome of such tests. Given results from, e.g., \citet{PengLilly10}, \citet{Behroozi13}, and K14, multiple paradigms of galaxy evolution would seem to (at least have the potential to) describe all of the data we have examined as well as ours. Those works outline two other perhaps equally effective yet largely incompatible frameworks operating at the descriptive level of the G13 model.

The first is the grow-and-quench paradigm against which we have principally contrasted our model \citep[e.g.,][]{PengLilly10}. O13 and \citet{Dressler16} demonstrate that there must be some galaxies that do not follow \MS-defined growth histories (Section \ref{sec:motivation}). However, if those exceptions are not seen as overly damaging, it seems likely that a model in which \MS\ or $\Mstel(\Mhalo)$ evolution reveals {\it governing principles} behind galaxy growth---halted by some agent uncorrelated with initial conditions---is compatible with everything we have so far explored. As such, we do not rule it out as viable description of galaxy evolution.

The second is the quasi-stochastic model of K14. In some aspects, this paradigm is quite similar to ours, yet in others it could not be more different.

The most notable similarities are (1) that key physics is encoded by the {\it scatter} in, e.g., the \MS; (2) that the cessation of star formation comes from the same phenomenon that sustains it (there is no explicit quenching); and (3) that controlling mechanisms must be inferred at a high level from the formalism of the model itself [in K14, the distribution of $\delta\sfr/\delta t$ updates ($\sigma$); in G13, the distribution of $(T_{0},\tau)$].

Yet, K14 reaches these conclusions by positing that most of galaxy evolution is governed by non-deterministic SFH ``updates'' to equilibrium conditions. These can be arbitrarily discontinuous and explicitly forbid parametrization. That is, the act of parametrizing the K14 SFHs---e.g., by a lognormal---destroys the ability to learn about the core physical process, which is whatever sets the spectrum of quasi-stochastic updates. K14 showed that these features make that model a good description of the histories of local dwarfs \citep[as directly measured by][]{Weisz14}, which our model does not contain, but likely also does not describe (Section \ref{sec:construction}). In these senses, it is {\it opposite} the G13 model.

That said, there is some suggestion that the spectrum of $\sigma$s in K14 is closely tied to initial halo conditions, specifically density (D.~Kelson, private communication). If so, the {\it prima facie} lack of a way to correlate/corral SFHs, e.g., spatially---an observed fact and something we believe is encoded by $(T_{0},\tau)$---may be alleviated, or at least is likely to derive from the same mechanism we are espousing. One can see how this might work: Where G13 suggests a galaxy is ``finished,'' K14 posits that it ``last updated a long time ago.'' Yet, both would agree, e.g., that passive (old) galaxies at any epoch should be denser than starforming (young) counterparts \citep{Martig09, Valentinuzzi10a, Fang13, Poggianti13a, Poggianti13b, Barro15, Morishita16b} because they reflect the denser nature of the Universe at the time they were last active.\footnote{G13 and K14 may be intimately related by the central limit theorem. K14 uses it explicitly to show that the shape and evolution of the low-mass \MS\ may reveal nothing but mathematical inevitabilities. G13 invokes the theorem implicitly---lognormals are the limit for random multiplicative processes---but, taking our ansatz seriously and letting K14 describe {\it parts} of galaxies, there might be a deep formal connection between the two models.}

The upshot is that descriptions in which galaxies:
\bitem
	\item evolve together along a physically informative \MS\ and are quenched \citep[e.g.,][]{PengLilly10};
	\item have diverse histories that naturally rise and fall as determined by initial conditions (G13);
	\item grow quasi-stochastically, such that the \MS\ reflects a foregone mathematical conclusion (K14);
\eitem
may be equally compatible with many key statistical observations in galaxy evolution. This suggests either that the data are unconstraining (see below), or that {\it formalism substantially influences physics}: If a model posits that galaxies evolve {\it along} the \MS\ [or $\langle\Mstel(\Mhalo)\rangle$], it cannot escape invoking what models with emergent \MS s (G13, K14) see as extraneous quenching prescriptions. It is must also neglect potentially revealing SFH diversity because it contains no formal resource but a description of the mean (without an ansatz for scatter).

At root, we are arguing that there is a meaningful lack of understanding (or at least consensus) regarding {\it how to describe} galaxy evolution (see \citealt{Taylor15} for extensive further discussion in the context of the $z\approx0$ stellar mass function). As such, even the most basic questions---How predictive of a galaxy's $\sfr(t)$ is its initial halo overdensity? Does the \MS\ describe an emergent or driving phenomenon? How far does stochastic growth take you?---are not well (or widely) understood. If so, such ignorance may supersede questions at the level of ``Is strangulation or AGN feedback the dominant quenching mechanism?''

This leads us to the third implication of our results.

\subsection{Epistemological Implications}
\label{sec:contentFree}

\subsubsection{Many Core Metrics Have Limited Discriminatory Power}

If nothing else, the results presented here, and in G13, \citet{Abramson15}, and \citet{Dressler16} suggest that a loosely constrained collection of SFHs governed by no explicit physical prescriptions does a fair job at describing many observations generally approached using more sophisticated techniques, such as semi-analytical modeling or hydrodynamic simulation. This is in no way to diminish the latter---they are the only routes to a first-principles explanation of the details of galaxy evolution. However, as very few such details {\it must be understood} to reproduce the data explored here, the latter would seem insufficient to confirm/falsify these models.

For example, at least at $z\lesssim2$, $\sigms$ has been shown to reflect: (1) Hubble timescale differentiation of the galaxy population (G13; this work); (2) burstiness on $10^{7}$ yr timescales \citep[e.g.,][]{Hopkins14}; and (3) fluctuations on arbitrary timescales (K14). If this metric accommodates {\it all possible timescales} for SFH variability, then it alone is a poor test of physical models with implications for that timescale.\footnote{Though future, e.g., deep spectroscopic observations at many $z$ may substantially increase the potency of this constraint by providing additional information on stellar populations (Section \ref{sec:betterTests}).}

Ultimately, we are suggesting that the data examined here and in many other works may constrain some basic mathematical properties of SFHs---e.g., that they rise and fall asymmetrically in time with independent half-mass-times and widths (Appendix \ref{sec:AA})---but not much else. To learn more, different information is needed.

\subsubsection{Towards More Discriminating Tests}
\label{sec:betterTests}

To progress, the dimensionality of the problem must be increased on both small and large scales.

Regarding the former, it may be that critical astrophysics are hyper-local \citep[e.g.,][]{Bigiel08}, so integrated quantities---$\sfr$, $\Mstel$, $Z$, $r_{e}$, etc.---are too low-resolution to be discriminating. Beyond diluting information, reducing galaxies to these numbers may cripple our ability to connect progenitors to descendants. Accounting for at least the fact that galaxies comprise bulges and disks changes the interpretation of the \MS\ \citep{Abramson14a}. It must also change assessments of which galaxies evolve into which others. Where else will similar moves be even more edifying?

The acquisition of high spatial-resolution IFU-like spectroscopy for even a few thousand galaxies over a fair range of cosmic time could be decisive. {\it JWST} and forthcoming extremely large ground-based telescopes (ELTs) will provide these data. Access to, e.g., gas/stellar metallicity {\it gradients} at many epochs will not only support much more sensitive tests of the phenomena that control gas processing---and thus galaxy growth---but may also provide better ways to link progenitors and descendants and thereby stitch together different cross-sectional samples. Both will greatly enhance our knowledge of the details of galaxy evolution.

Pilot surveys of this nature are underway---GLASS \citep{Treu15, Jones15, Vulcani15}, GASP (Poggianti et al., in preparation), {\sc Atlas3D} \citep{Cappellari11}, {\sc 3D-HST} \citep{Brammer12,Nelson15}, CALIFA \citep{Sanchez12}, {\sc MaNGA} \citep{Bundy15}, SAMI \citep{AllenSAMI15}, KROSS \citep{Magdis16}---and proving quite powerful. Hence, the ``dimensionality problem'' at small scales may itself soon be reduced.

Doing so would amount to winning most, but crucially not all of the battle. To achieve this, any physics inferred/ancestral mapping provided by {\it JWST} and the ELTs must be contextualized on {\it large scales} using clustering/``two-point'' statistics. This is because models must not only generate galaxies with the right $(\Mstel,\sfr)$, but also put them in the right places at the right times. There is strong theoretical and empirical evidence for correlations between baryonic and halo properties at single epochs \citep[e.g.,][]{Weinmann06, Guo10, Hearin13, Kawinwanichakij14, Tinker16c}. Truly viable models must also recover the {\it evolution} of these correlations \citep[e.g.,][]{Conroy06} over large dynamic ranges in $\Mstel$, $\sfr$, and time, thereby overcoming the substantial diversity in halo or stellar mass growth histories leading to the same $(\Mstel,\sfr)$ coordinate at any one moment (\citealt{Behroozi13a,Dressler16,Tinker16c}).

This will require deep, wide surveys that sample and tomographically track the growth of the full range of galaxy environments over much of time. {\it WFIRST} \citep{WFIRST_SDT_FINAL, Spergel15} will provide such data, enabling the clustering measurements needed to discriminate between scenarios that equally reproduce $\Mstel$ and $\sfr$ distributions, but are based on different mappings to underlying halo activity.

For example, the G13 framework implies that, at some early time, the galaxy correlation function must be biased towards systems of the {\it highest}---as opposed to lowest---$\sfr$s \citep[see also][]{Feldmann16}. If this is not seen, quite simply, the model is wrong. Additionally, for the same reason, our model suggests that the UV sources {\it observed} during reionization may differ substantially from those {\it powering} it: the former live in overdensities and are therefore mature, while the latter live in less-dense regions and are thus much younger \citep[and might have different spectra; ][]{Furlanetto04,FZH04,Wise14,Davies15,Stark16}. This need not be the case in \MS-driven or quasi-stochastic models, where environmental effects either might not exist or might be washed out. Hence, high-$z$ observations sampling a high dynamic range of environments should enable powerful tests of these paradigms (see \citealt{Mirocha16} for quantitative explorations in this vein).


\section{Summary}
\label{sec:summary}

We have shown that the following observational metrics of galaxy evolution:

\benum
	\item the evolution of $\langle\ssfr\rangle$ for low-mass ($\log\Mstel\in[9.4,10]$) galaxies since $z=7$ (Section \ref{sec:meanSSFR}, Figure \ref{fig:meanSSFR});
	\item the evolution of the galaxy stellar mass function since $z=8$ (Section \ref{sec:SMFs}; Figure \ref{fig:SMFs});
	\item the evolution of the slope of the $\sfr$--$\Mstel$ relation (the SF ``Main Sequence''; \MS) since $z=6$ (Section \ref{sec:MSslope}, Figure \ref{fig:SFMSslope});
	\item the transition from ``fast-'' to ``slow-track'' quenching over cosmic time (Section \ref{sec:GV}, Figure \ref{fig:GVtransit});
	\item galaxy downsizing (Section \ref{sec:downsizing}, Figure \ref{fig:medTtau});
	\item the bulge mass fractions of local galaxies as a function of $\sfr/\Mstel$ or $\sfr/\Mdisk$ (Section \ref{sec:BT}, Figures \ref{fig:BTstuff}, \ref{fig:prediction});
\eenum
are naturally and well reproduced by a purely mathematical, explicitly quenching-free description introduced in \citet[][``G13'']{Gladders13b} {\it that was not designed to match any of them}. In this model, galaxies have continuous, smooth, lognormal star formation histories (SFHs) constrained mainly by the evolution of the cosmic $\sfr$ density (Section \ref{sec:construction}; Figures \ref{fig:fit}, \ref{fig:schema}). The cessation of star formation is thus linked strictly and identically to the same process causing SFHs to generically transition from rising to falling states, suggesting that---for most galaxies---no additional terminating process is necessary \citep[see also][]{AragonCalvo16}.

\bitem
	\item The model's emergent interpretation of the \MS\ scatter as {\it Hubble timescale diversification} of galaxy SFHs combined with the lognormal's ability to rise and fall faster than this locus enable its successes (Section \ref{sec:genotype}; Figures \ref{fig:todaySFMS}, \ref{fig:SFMStrends}). They allow G13 to recast binary ``starforming'' and ``non-starforming'' labels as a spectrum of {\it extended} and {\it compressed} SFHs, none of which must be subject to qualitatively different physics (Section \ref{sec:GV}; Figure \ref{fig:GVtransit}), or any beyond that which causes SFHs to turn over. This implies: 
	\benum
		\item[i.] Progenitors of galaxies within a factor of 2 of the Milky Way's current stellar mass may span a factor of $\sim$30 at $z=3$; i.e., galaxies do not evolve ``along'' the \MS, but ``through'' it (Section \ref{sec:moreTau}, Figures \ref{fig:delSFMS}, \ref{fig:delTau}). 
		\item [ii.] If so, the utility of that scaling relation in characterizing evolutionary trajectories or gleaning physics is dubious: the mean SFH$(\Mstel,\sfr;\, t)$ is unrepresentative of individual SFHs and insensitive to potentially critical mechanisms that differentially accelerate galaxy growth.
		\item[iii.] Trends displayed by (series of) single-epoch/cross-sectional data may be {\it orthogonal} to true evolutionary/longitudinal trends and may therefore support misleading inferences. 
		\item [iv.] Star formation {\it timescales}, $\tau$, present a key manifestation of this issue, appearing uncorrelated with $\sfr$s at fixed $\Mstel$ when examining the full galaxy population at $z\lesssim2$, but showing strong trends once systems cohabiting the \MS\ with true evolutionary cohorts but destined for different final masses are removed [Sections \ref{sec:newSpaces}, \ref{sec:toPhysics}; cf.\ Figures \ref{fig:SFMStrends} ({\it bottom}), \ref{fig:delSFMS}]. 
		\item [v.] We interpret (iv) to mean that understanding whatever sets the {\it width} of (lognormal) SFHs is critical to placing observations in the context of a comprehensive narrative of galaxy evolution. 
	\eenum
	\item Towards that end, we posit that $\tau$ is closely tied to baryonic surface densities, gas consumption timescales, or ``compaction'' phenomena based on highly suggestive comparisons between trends found here and those in the literature (Section \ref{sec:toPhysics}, \ref{sec:BT}; Figures \ref{fig:delTau}, \ref{fig:BTstuff}, \ref{fig:prediction}).
	\item Since G13 fixes $\tau$ once for each SFH for all time, we infer from the above that {\it initial conditions are powerfully predictive of a galaxy's SFH}, and can serve as a broader organizing principle in which quenching---here, a Hubble-timescale phenomenon---might be more instructively couched (Section \ref{sec:ICs}). We suggest the baryonic overdensity in some optimal aperture at $z>8$ (the G13 start time) as a potential characterization of those conditions.
	\item If so, in many instances, there need not be a distinction between environmental and internal mechanisms for quenching, which is better set as the ``peeling off'' of the densest tail of the starforming population at any epoch. Internal galactic (gas) densities correlate with those on super-galactic scales, leading to spatially confined cohorts of equally rapidly evolving, overdense (i.e., bulge-dominated) galaxies that all ``finish'' their SFHs before their less dense (i.e., diskier) counterparts in lower-density regions (Section \ref{sec:natureIsNurture}). This scenario seems naturally linked to the core \LCDM\ concept of ``assembly bias,'' and echoes the ideas of \citet{Holmberg64}, \citet{Tinsley68}, and \citet[][]{Dressler80}.
\eitem

We contend that all of the above makes G13 a descriptively powerful and physically informative paradigm of galaxy evolution that is meaningfully different from the dominant ``grow-and-quench'' interpretation. However, we acknowledge that models set in the latter framework---and at least one other based on {\it opposite} assumptions to our own \citep{Kelson14}---can reasonably claim to match some or all of the observations examined here as well as G13 (Section \ref{sec:descriptions}). As such, independent of G13's ultimate accuracy, we also contend that an important lack of consensus exists regarding basic aspects of what a ``good'' description of galaxy evolution looks like. These would stymie any attempt to set physical phenomena in a broader unifying explanatory theory.

We attribute much of this ambiguity to a lack of discriminating power in current data (Section \ref{sec:contentFree}). This may be due to the requirement to project galaxies onto planes defined by spatially-/temporally-integrated quantities. Such issues will be remedied by the next generation of telescopes---{\it JWST}, {\it WFIRST}, and thirty-meter class ground-based facilities---which will enable us to better connect progenitors to descendants based on spatially resolved spectroscopy over large swaths of cosmic time, contextualize those findings in diverse environments, and therefore better assess both the physics driving and the story of galaxy evolution.\\


\section*{}
L.E.A.\ thanks Dan Kelson, Jordan Mirocha, and Lindsay Young for intense, informative, and inspiring conversations. He also thanks Sean Johnson for software contributions; Josh Speagle for edifying feedback; Tommaso Treu for his patience and guidance during the completion of this work; and Hsiao-Wen Chen, Andrey Kravtsov, and Rich Kron for their helpful comments during its inception. Finally, he thanks Barbara Zera Abramson for valuable editorial and organizational assistance, and the anonymous referee for his/her insightful remarks. B.V.\ acknowledges support from an Australian Research Council Discovery Early Career Researcher Award (PD0028506).


\clearpage
\bibliographystyle{apj}
\bibliography{/Users/labramson/lit.bib}



\appendix

\section{A: Axiomatic Underpinnings of G13}
\label{sec:AA}

\subsection{Defining the G13 Class}

A key aim of this paper is to explore how axioms of SFH models affect physical interpretations of data (Section \ref{sec:discussion}). The vehicle for this exploration has been a model presented in \citet{Gladders13b} that is sufficiently descriptive of the data to make this exercise worthwhile (Section \ref{sec:phenotype}). Yet, it is just one member of a class of models representing instantiations of similar ideas. We are therefore interested in the potential of the G13 {\it model class} beyond the specific realization published in the above text. To clarify what this means, we must clarify what the G13 model class is. 

\citet[][]{Gladders13b} describes a ``species'' of the model family defined by the following traits:
\benum
	\item The basic modeling unit is the individual SFH as assigned to each member of an input set of (real) galaxies;
	\item The SFHs are continuous and smooth;
	\item The SFHs rise and fall;
	\item The SFHs are constrained mainly in the ensemble (e.g., by the time derivative of their {\it sum}).
\eenum
In terms of other models mentioned in-text, (1) is unique to this family; (2) is shared by quenching-based/\MS-driven models (up to the quenching event; e.g., \citealt{PengLilly10}); and (4) is shared by the diversification-based stochastic model of \citet{Kelson14}. Trait (3) {\it appears} in all, but is not required by K14, and SFHs that fall faster than the \MS\ (see below) cannot be produced by \MS-based approaches, driving their requirement for additional quenching mechanisms. Thus, though akin to a ``basis-set'' in ``paradigm-space,'' these perspectives are not quite orthogonal.

Trait (1) sets the basic aspect of the G13 family: a longitudinal survey of the input galaxy sample. This point is important and the subject of Appendix \ref{sec:AB}. We do not know if the G13 instantiation {\it specifically} (see below) is a singularly good representative of this family, but we have shown that at least one ``cousin''---a collection of Gaussians---is not viable (see G13). 

\begin{figure*}[t!]
\figurenum{A1}
\centering
\includegraphics[width = 0.95\linewidth, trim = 1cm 0cm 1cm 0cm]{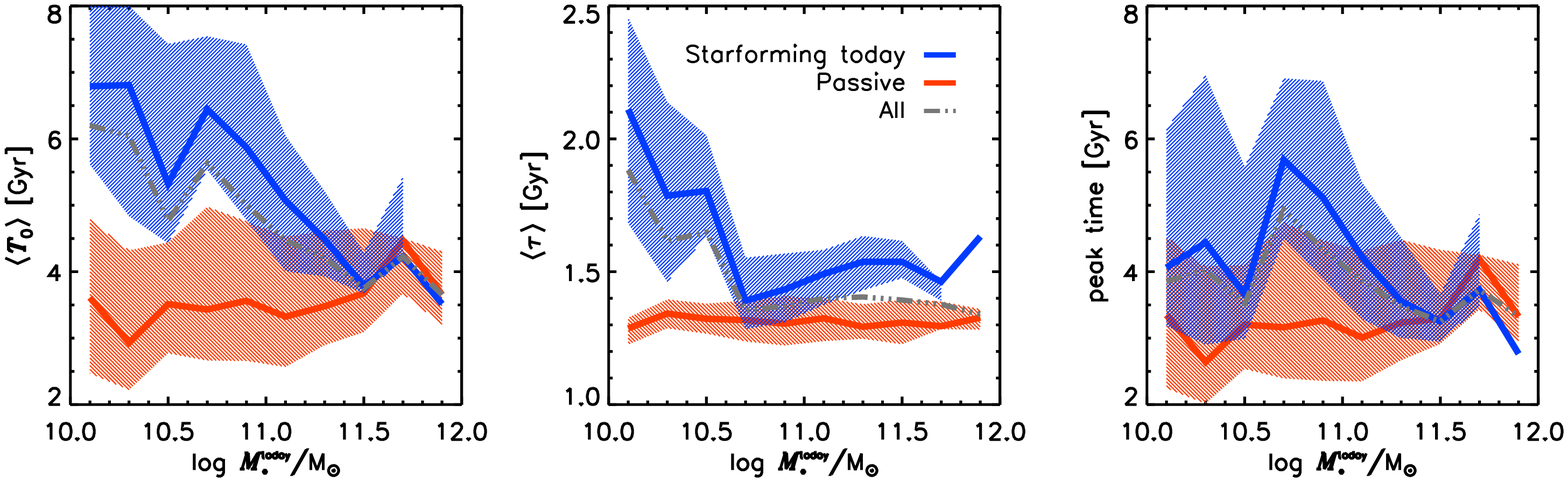}
\caption{{\it From left}: median $T_{0}$--, $\tau$--, and $t_{\rm peak}$--$\Mstel^{\rm obs}$ trends as output by G13 for ``starforming'' (blue) and ``non-starforming'' (orange) present-day galaxies (split at $\log\ssfr=-11$). Band widths denote 25\%--75\% parameter spreads; grey dot-dashed lines show trends for all galaxies. These quantities show different mass and population dependencies. That $\langle T_{0}\rangle$ monotonically decreases while $\langle\tau\rangle$ stays flat at $\log\Mstel\gtrsim10.5$ for today's starforming systems demonstrates the independence of these parameters [trait (6)] needed to produce the large-$T_{0}$, small-$\tau$ objects identified by O13. Note that the decorrelation of all quantities with $\Mstel$ for the passive population reflects the same lack of $\ssfr$ constraints discussed in the context of the {\it UVJ} diagram and downsizing (Sections \ref{sec:UVJ}, \ref{sec:downsizing}; Figures \ref{fig:UVJ}, \ref{fig:medTtau}). The apparent bimodality in $\langle\tau\rangle$ at $\log\Mstel\sim10$ may also be driven in part by this issue, but much of it may not (Appendix \ref{sec:AC}), demonstrating that population bifurcation can emerge from from models where SFHs are drawn from a unimodal parent distribution (Figure \ref{fig:bimodality}).}
\label{fig:TtauMass}
\end{figure*}

Extrapolating from this and the results of O13, the {\it class} of viable models to which G13 belongs exhibits two traits beyond those listed above:
\benum
	\item[5.] The SFHs are {\it time-asymmetric}, such that they rise faster than they fall;
	\item[6.] The SFHs are characterized by at least two parameters, such that their half-mass-times are formally independent of their widths (Figure \ref{fig:TtauMass}).
\eenum
The ability of this class to produce the late-peaking, narrow SFHs demanded by O13 was a result of (6). This is a consequence, however, of the more basic fact that (6) allows SFHs to fall faster than the \MS---i.e., {\it to exit the starforming population} without invoking any mechanism beyond whatever causes SFHs to decline, generally. As such, (6) is {\it also} the reason G13 can avoid explicit quenching (see discussion of Figure \ref{fig:todaySFMS} in Section \ref{sec:sigms}). Yet, because Gaussian SFHs exhibit this trait and are {\it not} good descriptions, this criterion alone is insufficient. The lognormal fulfills both (5) and (6), though it was not selected for this reason.

The G13 paradigm would be invalidated if the following {\it class-level} failures were identified. From most to least damaging:
\bitem
	\item Large, abrupt discontinuities are required to explain a significant fraction of SFHs for systems dominating the Universe's mass and star formation budget over a significant fraction of cosmic time (smooth, continuous parameterizations cannot describe most SFHs);
	\item There is no interesting physical interpretation of (the independence of) SFH half-mass times and widths;
	\item There exists no physical quantity at, e.g., $z\gg8$ to which the parameters of smooth, continuous SFHs could be tied that correlates meaningfully with observed properties at, e.g., $z\ll8$.
\eitem
Such findings would demonstrate flaws at the core of the G13 modeling framework and cause us to abandon it.

\subsection{Defining the G13 Instantiation}
Though nominally central to the model---and perhaps ultimately physically meaningful---the lognormal form is really just one of a suite of mathematical details that distinguish G13 {\it as currently realized} as an instantiation within the class just described. Its defining structures are displayed below.

The lognormal SFH of the $i$-th galaxy is given by:
\beq
	\sfr(t)_{i}\propto\frac{\exp\left[-\frac{(\ln t - T_{0,\,i})^{2}}{2\tau_{i}^{2}}\right]}{t\sqrt{2\pi\tau_{i}^{2}}},
\label{eqn:logSFH}
\eeq
under the boundary conditions:
\beq
	f\int^{t^{\rm obs}}_{t_{\rm 0}}\sfr(t)_{i}\,dt = M_{\ast,\,i}^{\rm obs},
\label{eqn:massCons}
\eeq

\beq
	\sfr(t^{\rm obs})_{i} = 
		\begin{dcases}
			\sfr_{i}^{\rm obs} & {\rm if}~\sfr_{i}^{\rm obs} > {\rm thresh} \\
			(-\infty,~{\rm thresh}]		 & {\rm otherwise},
		\end{dcases}
\label{eqn:sfrCons}
\eeq
and
\beq
	\frac{1}{V} \sum_{i=1}^{N}\sfr(t)_{i} = {\rm SFRD}(t).
\label{eqn:sfrdCons}
\eeq
That is, the current G13 model is a set of $N(=2094)$ half-mass-time/width parameter pairs, $\{(T_{0},\tau)_{i}\}$, that ensure (1) the $i$-th galaxy's lognormal SFH (Equation \ref{eqn:logSFH}) leads to its observed $(\Mstel,\sfr)^{\rm obs}$ at the correct epoch (\ref{eqn:massCons}, \ref{eqn:sfrCons}); and (2) the ensemble of SFHs sums to the correct SFRD at all times (\ref{eqn:sfrdCons}; $V$ is the comoving volume of the input sample). We assume that SFHs for galaxies with $\sfr^{\rm obs}$ upper-limits can take any value below that threshold at $t^{\rm obs}$ ($0.05~\Msun~{\rm yr^{-1}}$; G13). We also adopt a constant IMF-dependent $\sfr\mapsto\dot\Mstel$ conversion factor, $f\sim0.7$ (Salpeter). Graphically, Figures \ref{fig:schema} and \ref{fig:todaySFMS} depict these equations.

\begin{figure}[h!]
\figurenum{A2}
\centering
\includegraphics[width = 0.6\linewidth, trim = 1cm 0cm 1cm 0cm]{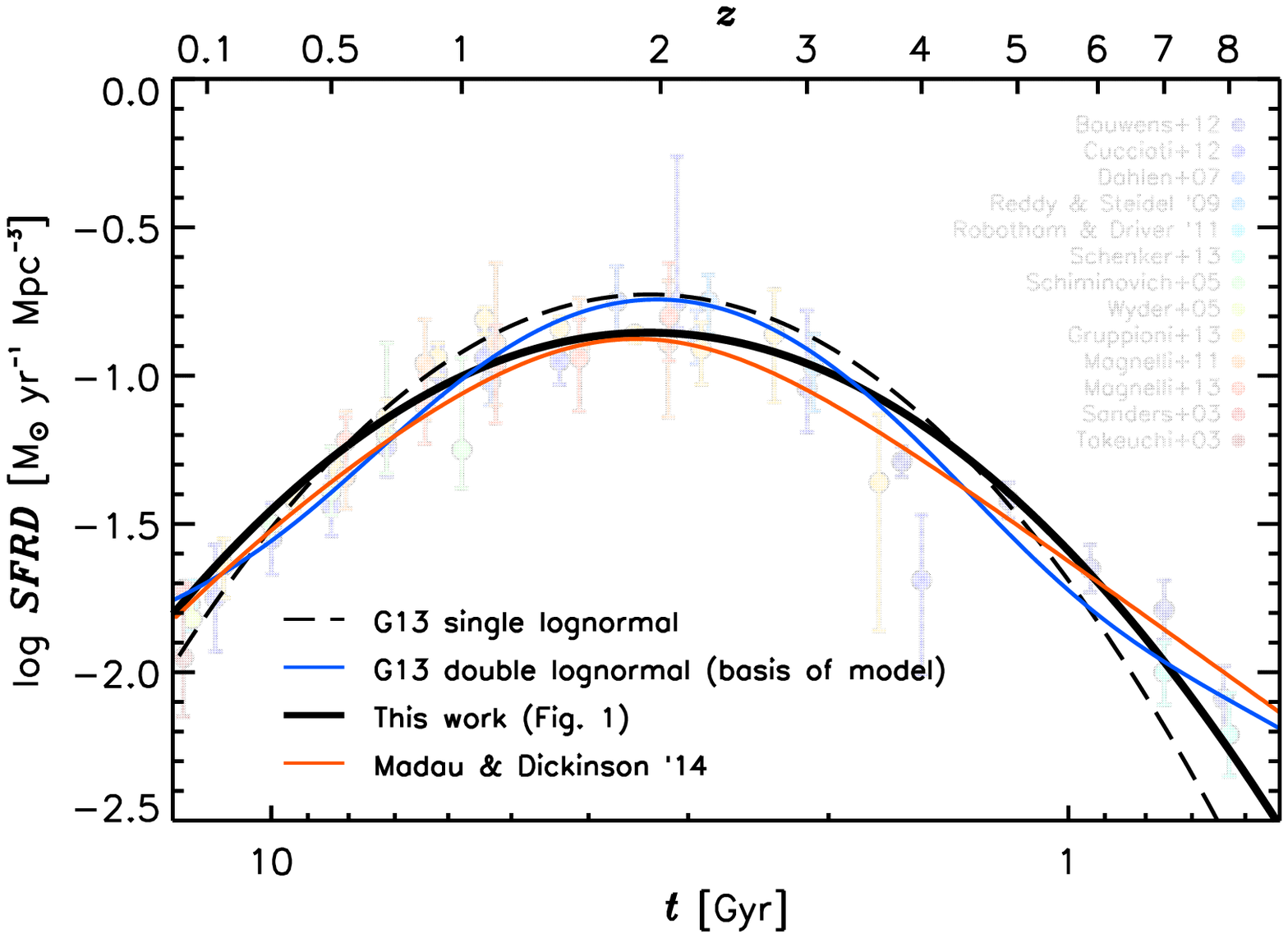}
\caption{Parameterizations of the cosmic SFRD$(t)$. {\it Solid black line}: new best-fit single lognormal from Figure \ref{fig:fit}. {\it Dashed line and solid blue line}: the single- and double-lognormal fits from the original G13 paper, \resp. The former $[(T_{0},\tau)_{\rm Uni} = (1.54, 0.57)]$ was the inspiration for that work; the latter $[(T_{0,1}, T_{0,2}; \tau_{1}, \tau_{2})_{\rm Uni} = (1.39, 2.80; 0.46, 1.19)]$ constrained the SFHs analyzed here \citep[and in][]{Abramson15}. {\it Orange line}: the best-fit double power-law from \citet{MadauDickinson14}. The fits are similarly good, though differences at the highest-$z$ may affect, e.g., the evolution of the stellar mass function (Figures \ref{fig:SMFs}, \ref{fig:SMFincompl}), and {\it UVJ} diagram (Figure \ref{fig:UVJ}).}
\label{fig:manySFRDs}
\end{figure}

The original treatment also entailed a merger prescription (see G13), but such events were neglected in the calculations used here. Given results from \citet{Leitner12}, \citet{Behroozi13}, \citet{Abramson15}, and our own numerical experiments, this appears safe, though it may impact number counts at $z\gtrsim2.5$ and $\log\Mstel<10$ (Figures \ref{fig:SMFs}, \ref{fig:SMFincompl}; cf.\ \citealt{Abramson15}, Figure 1) and likely implies that G13 SFHs represent mean histories for ensembles of merging sub-systems at similar or higher $z$.

Various G13 ``siblings'' are characterized by lognormal or double-lognormal forms for SFRD$(t)$ (Figures \ref{fig:fit}, \ref{fig:manySFRDs}), and the use of $\ssfr$ distributions at $\langle z\rangle\simeq0.3$, 0.5, 0.7, and 0.9 to further constrain $\{(T_{0},\tau)_{i}\}$. The results in this paper were derived from the maximally constrained (double-lognormal SFRD + sSFR distributions) realization, which is qualitatively similar to the others, though likely differs in quantitative detail. Models generated using a different SFRD form and/or a different SFH parameterization may ultimately be shown to exhibit cross-sectional projections (Section \ref{sec:glossary}, Appendix \ref{sec:AB}) so closely aligned with G13's as to be indistinguishable. Conversely, we might find that imposing additional (e.g.) $\ssfr$ constraints on the current model cannot remedy the ``superficial'' issues mentioned in-text (Section \ref{sec:phenotype}). In either case, we will learn something important about the nature of smooth (lognormal) SFHs, or identify regimes where they are inappropriate. We are obtaining more discriminating data to enable such tests of how the G13 SFHs relate to those of individual galaxies (\citealt{Dressler16}, Oemler et al., in preparation).

\section{B: Further Epistemological Implications of G13}
\label{sec:AB}

The G13 family comprises theoretical longitudinal surveys of real galaxy samples. This fact formally limits any such model to describing {\it only} true (theoretical) progenitors/descendants. G13 is therefore subject to a different kind of incompleteness than that typically discussed in astronomy. 

A sample is ``complete'' in the usual sense if it fills a representative volume of $(\Mstel, \sfr, ...)$ space {\it at the epoch from which it is drawn}. This is ``cross-sectional'' completeness, and investigators go to great lengths to compile and compare such samples at multiple epochs. 

G13 does not describe these data. It does not produce a time-ordered series of complete cross-sectional samples, but {\it cross-sectional views of a longitudinally complete model} (i.e., one containing the entire set of progenitors of {\it a single} cross-sectional sample). By comparing these two distinct entities, this work has perpetrated a subtle but significant sleight-of-hand. Because there {\it should} be objects in the data that {\it do not represent} progenitors of our input sample---and so {\it should not} be described by the model's projections---any such comparisons may have been inappropriate. 

Yet, the fact that the G13 model performs as well as it does in this (formally dubious) exercise suggests otherwise: In many contexts, none of the above seems to matter. One context where it might is the (non-)evolution of the scatter in the \MS\ ($\sigms$), but because we believe we can apply additional constraints to match this using the current G13 approach, even this might not be the case (up to some ultimate redshift limit; Section \ref{sec:meanSSFR}). If so, it would seem that a complete set of galaxies (in the usual, cross-sectional sense) at one epoch corresponds to a complete set of progenitors at all epochs, at least over interesting areas of parameter space (see, e.g., Figure \ref{fig:SMFincompl}). 

Note: {\it this does not mean that cross-sectional trends are equivalent to evolutionary trends}, just that complete cross-sectional data at earlier epochs contain/represent analogs of many/all of the progenitors of complete cross-sectional samples at later epochs. Proving (under which conditions) this statement is true would be a boon to the science.

\section{C: From One, Many: Bimodality and the $(T_{0},\tau;\, \lowercase{t})$ Continuum}
\label{sec:AC}

\begin{figure}[h!]
\figurenum{C1}
\centering
\includegraphics[width = 0.75\linewidth, trim = 1cm 0cm 1cm 1cm]{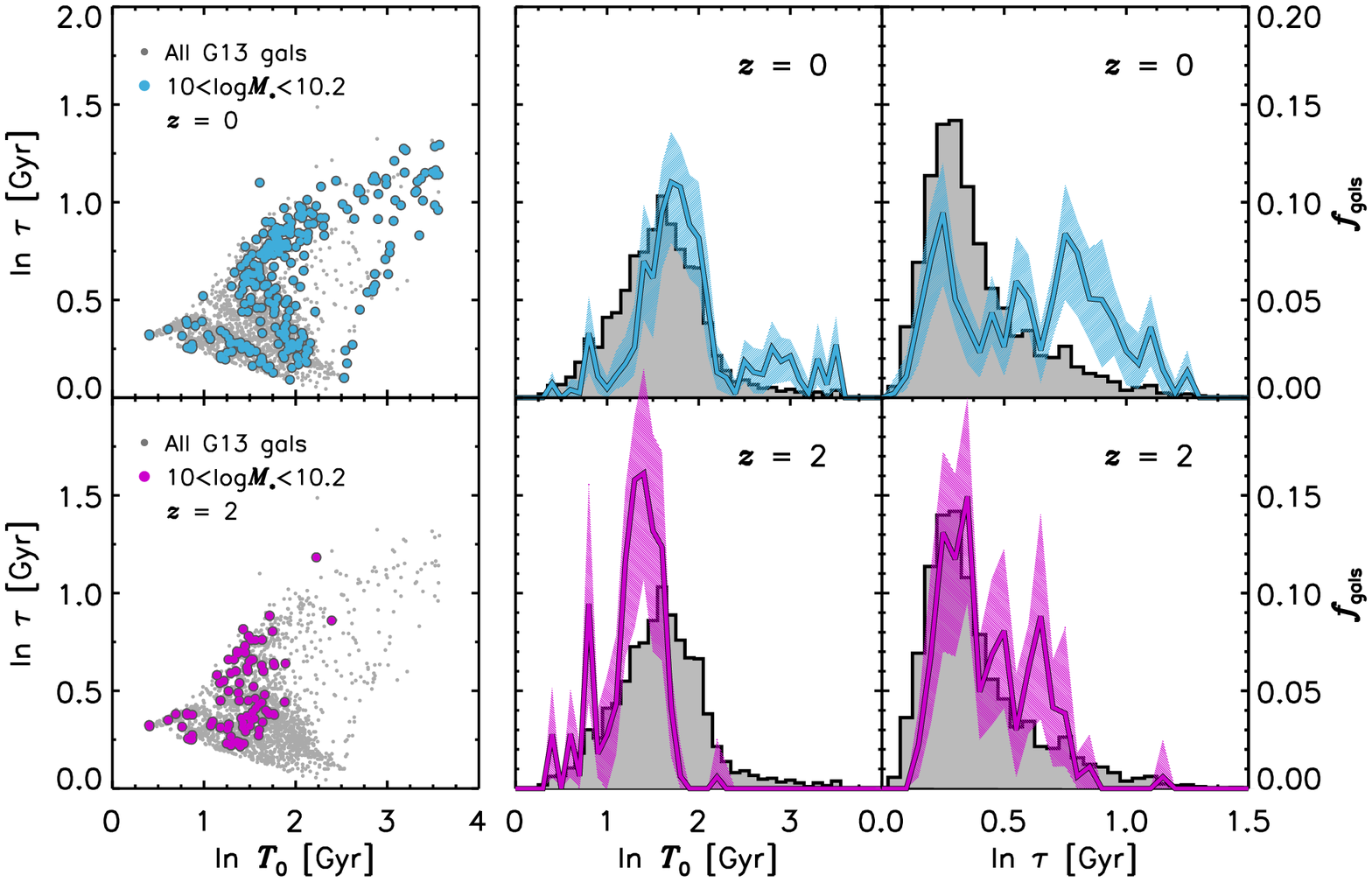}
\caption{{\it Top, from left}: the joint $(T_{0},\tau)$, and 1D $T_0$ and $\tau$ distributions for all G13 SFHs (grey points/histograms; shading is 95\% credibility) and those with $z=0$ $\log\Mstel\sim10$ (blue). {\it Bottom}: the same, but at $z=2$ (purple). A $\tau$ bimodality emerges at $z=0$ as anticipated from Figure \ref{fig:TtauMass}, but not at $z=2$, illustrating how similar $(\Mstel,\sfr)$ regimes correspond to different $(T_{0},\tau)$ regimes different times. As a consequence, {\it contemporaneous}, equal-$\Mstel$ passive and active galaxies will also lie in different parts of $(T_{0},\tau)$ space: the former in regions that corresponded to ``starforming'' long ago, the latter in those that do so at the time of observation. This naturally allows such classes (defined at a single epoch) to bifurcate in $(T_{0},\tau)$, though they all hail from a unimodal parent distribution.}
\label{fig:bimodality}
\end{figure}

Figure \ref{fig:TtauMass} suggests a bimodal $(T_{0},\tau)$ distribution for today's low-mass starforming and non-starforming galaxies. This would seem to contradict our central claim that quenching is not a ``special'' process, but corresponds to the declining side of {\it all} SFHs, which are drawn from a $(T_{0}, \tau)$ continuum (see also Section \ref{sec:GV}). Figure \ref{fig:bimodality} illustrates why this is not the case.

From left to right, this Figure shows the joint $(T_{0},\tau)$, and 1D $T_{0}$ and  $\tau$ distributions for all G13 SFHs (grey), and those with $z\sim0$ ({\it top}) and $z\sim2$ ({\it bottom}) $10\leq\log\Mstel\leq10.2$ (colors). At either redshift, $T_{0}$ shows no bimodality, but at $z\sim0$, there is some in $\tau$ as anticipated and reflected in the joint distribution. However, this signal vanishes at $z\sim2$, belying a key point: Depending on the epoch, any SFH derived from a coordinate in the $(T_{0},\tau)$ continuum will be above or below a given $\ssfr$ threshold and therefore appear as a starforming or non-starforming galaxy (cf.\ lines of constant $\ssfr$ in G13, Figure 9, with those in \citealt{Dressler16}, Figure 10). That is, the SFRD is supported by different parts of $(T_{0},\tau)$ space at different redshifts.

Comparing the purple to blue distributions in Figure \ref{fig:bimodality} shows this: Galaxies of a given mass (and starforming class) occupy different locations in $(T_{0},\tau)$ space at different times. Hence, at any given time, starforming and non-starforming galaxies will, by definition, lie in different parts of the plane: one is coming into dominance, the other is (far) removed from it. Indeed, at early times, some portion of the plane represents SFHs that mathematically cannot achieve a given $(\Mstel,\sfr)$ state and will therefore be unpopulated (lower-left panel). Conversely, at late times, much more $(T_{0},\tau)$ space can lead to a given $(\Mstel,\sfr)$---there is more time for slowly evolving SFHs to mature---more-fully populating the diagram (upper-left panel). Hence, the age of the Universe inevitably influences which portions of $(T_{0},\tau)$ space dominate which portions of $(\Mstel,\sfr)$ space. This can naturally lead to the appearance of bimodality even though all systems were actually drawn from the same unimodal parent distribution.

A further practical consideration likely exacerbates any such natural segregation. As with all of our results (see Section \ref{sec:results}), the details of G13's description of $(T_{0},\tau)$ space depend on the data used to constrain the model. Since we used only $\ssfr$ distributions and the SFRD---and thus have no handle on {\it when} a galaxy quenched provided it was long-enough ago to lie below the $z\sim0$ $\sfr$ threshold (Equation \ref{eqn:sfrCons})---there can be significant separation between the passive (i.e., once-starforming) population, and that with SFRs currently required to support the SFRD: SFHs with no $z\sim0$ $\ssfr$ constraint tend towards the SFRD (Section \ref{sec:downsizing}, Figure \ref{fig:medTtau}), which peaks sufficiently early for those histories to finish while leaving plenty of time before present-day $\log\Mstel\sim10$ starforming galaxies must become important to the SFRD. The model has no mechanism to encourage it to fill-in this space, which would also contribute to the bifurcation discussed above.

\section{D: A Note on the Galaxy Stellar Mass Function}
\label{sec:AD}

\begin{figure}[h!]
\figurenum{D1}
\centering
\includegraphics[width = 0.8\linewidth, trim = 1cm 1cm 1cm 1cm]{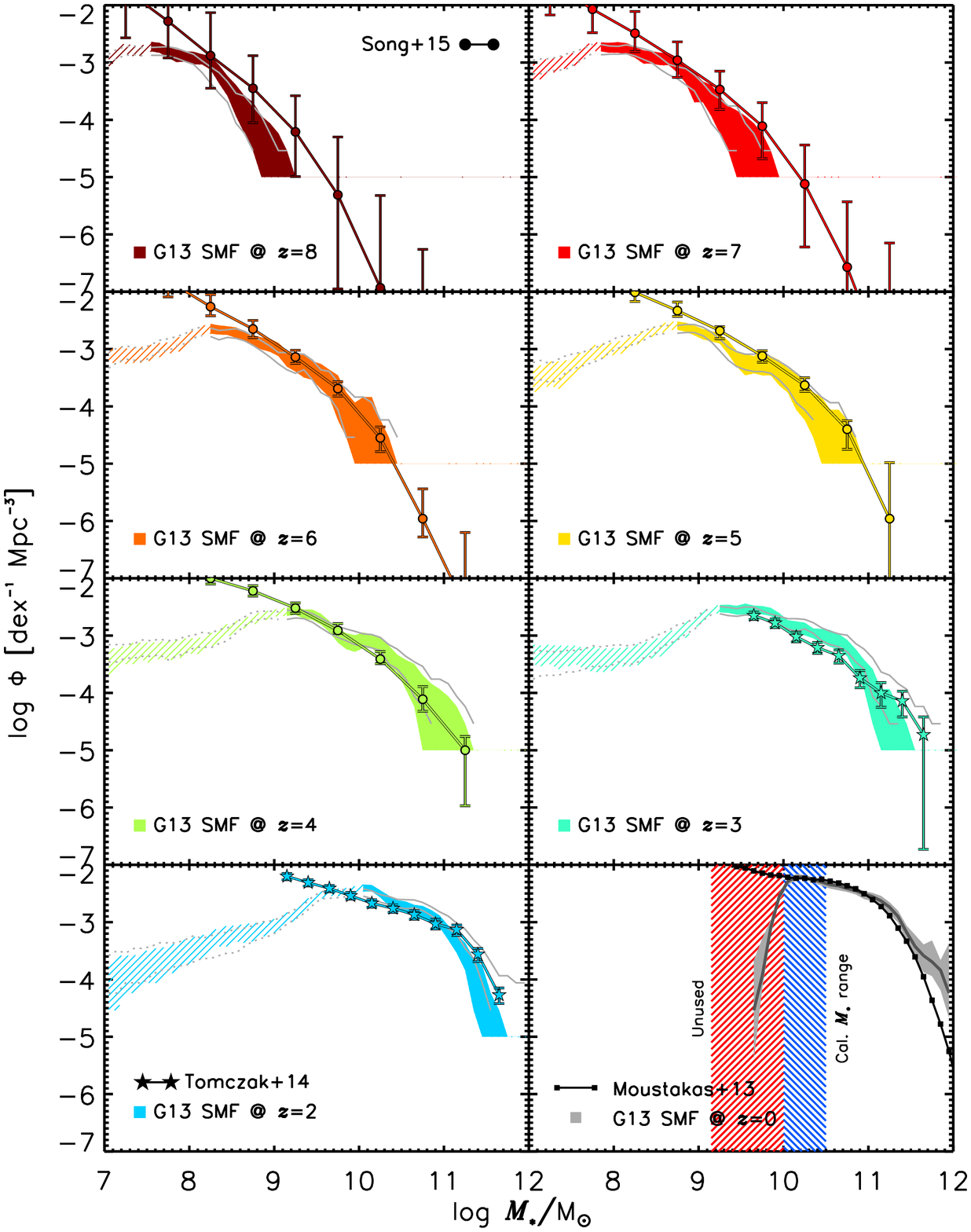}
\caption{G13 SMFs (Figure \ref{fig:SMFs}) broken-out by redshift to better illustrate cross-sectional incompleteness (hashed regions) and the effect of including the mild overabundance of massive galaxies in the G13 input sample compared to the full SDSS (\citealt{Moustakas13}; {\it bottom-right}; tracebacks shown by solid/dotted grey lines in other panels). Colors, data points, and filled bands are those in Figure \ref{fig:SMFs}.}
\label{fig:SMFincompl}
\end{figure}

For the sake of clarity, incomplete $\Mstel$ regimes (in the usual, cross-sectional sense) were omitted in Figure \ref{fig:SMFs}, showing the G13 predicted SMF evolution. We show the full G13 SMFs in Figure \ref{fig:SMFincompl}. Color-coding corresponds with the in-text Figure.

Also shown is the $z\approx0$ data \citep[][converted to a Salpeter IMF]{Moustakas13} used to calibrate G13 to an absolute scale ({\it bottom-right}). The grey curve shows the histogram of the input galaxy masses that serve as boundary conditions for the SFHs (Equation \ref{eqn:massCons}). The G13 sample evidently contains an overabundance of galaxies at the highest masses compared to the full SDSS data, representative of the general population of the local Universe. 

To ensure tracing-back a fair galaxy sample, the G13 input data were randomly resampled ($100\times$) to match the \citet{Moustakas13} SMF before deriving results at other epochs (colored bands). If the input data are not resampled, one obtains the light grey lines in each panel. Changes are not dramatic, though tension is eased slightly in some instances and increased in others.

\clearpage
\end{document}